\documentclass[
reprint, superscriptaddress,
amsmath,amssymb,
aps
]{revtex4-2}

\usepackage{graphicx}
\usepackage{siunitx}  
\usepackage{amsmath}
\usepackage{tabularx}
\usepackage{braket}   
\usepackage{comment}     
\usepackage[dvipsnames]{xcolor}
\usepackage[normalem]{ulem}
\usepackage{multirow}
\usepackage{float}

\newcommand{\be}{\begin{equation}}
\newcommand{\ee}{\end{equation}}

\date{\today}

\begin{document}

\title{
Anomalous Frequency Shift in Low-Loss Superconducting Granular Aluminum Resonators
}

\author{Patrick Winkel}

\email{patrick.winkel@alice-bob.com}
\email{Present address: Alice \& Bob, Paris, France}
\affiliation{Departments of Applied Physics and Physics, Yale University, New Haven, CT, USA}
\affiliation{Yale Quantum Institute, Yale University, New Haven, CT, USA}

\author{Daniil S. Antonenko}
\affiliation{Departments of Applied Physics and Physics, Yale University, New Haven, CT, USA}

\author{Vishakha Gupta}
\affiliation{Departments of Applied Physics and Physics, Yale University, New Haven, CT, USA}
\affiliation{Yale Quantum Institute, Yale University, New Haven, CT, USA}

\author{Neel Thakur}
\affiliation{Departments of Applied Physics and Physics, Yale University, New Haven, CT, USA}
\affiliation{Yale Quantum Institute, Yale University, New Haven, CT, USA}

\author{Pavel Kurilovich}
\affiliation{Departments of Applied Physics and Physics, Yale University, New Haven, CT, USA}
\affiliation{Yale Quantum Institute, Yale University, New Haven, CT, USA}

\author{Luigi Frunzio}
\affiliation{Departments of Applied Physics and Physics, Yale University, New Haven, CT, USA}
\affiliation{Yale Quantum Institute, Yale University, New Haven, CT, USA}

\author{Leonid I. Glazman}
\affiliation{Departments of Applied Physics and Physics, Yale University, New Haven, CT, USA}
\affiliation{Yale Quantum Institute, Yale University, New Haven, CT, USA}

\author{Robert J. Schoelkopf}
\email{robert.schoelkopf@yale.edu}
\affiliation{Departments of Applied Physics and Physics, Yale University, New Haven, CT, USA}
\affiliation{Yale Quantum Institute, Yale University, New Haven, CT, USA}

\begin{abstract}
Superconducting high-kinetic inductance materials like granular aluminum (grAl) are a versatile part of the circuit quantum electrodynamics (cQED) toolbox, provided their losses at microwave frequencies are low enough. To further advance the use of grAl in quantum devices, it is indispensable to identify the dominant loss mechanisms. The standard approach pairs electromagnetic simulations of resonator geometry with measured temperature and power dependence of resonance frequency and loss rate. Most materials follow the phenomenological theory of two-level systems (TLSs) at low temperatures, resulting in an initial decrease of the resonance frequency with increasing temperature. In our work, we observe an opposite behavior at the lowest temperatures: The resonance frequency initially increases with both temperature and readout power, contradicting the standard TLS model predictions. We are able to explain a part of the data by an alternative mechanism associated with the effect of superconducting quasiparticles in a granular system with spatially-nonuniform gap. Yet, the observed change in the resonance frequency with temperature is not matched by a proportional change in the loss rate.
We also observe anomalous positive frequency responses following high-energy events, characterized by several time scales. While anomalous behavior has been reported previously in grAl, the disagreement with standard models is particularly visible in our devices thanks to their exceptionally low loss rates.   
\end{abstract}

\maketitle

\section{Introduction}
Superconducting materials with high kinetic inductance have attracted considerable interest for quantum device applications, owing to a combination of favorable properties: their high sheet inductance enables compact inductor designs~\cite{Rotzinger_2017,Dupre_2017,IoanPop-2018,Niepce_2019,zhang_microresonators_2019,Winkel_grAl_transmon, Glezer_2020, rieger_granular_2023, kristen_random_2023, Frasca_2023_lin,Gyenis-experiment,Charpentier2025}, while they typically also offer strong resilience to magnetic fields~\cite{ borisov_superconducting_2020,Xu_2023,Khalifa_2023, Zapata_2024,Frasca_2024, Roy_NbN_2025,Janik_2025} and low non-linearity\,\cite{maleeva_circuit_2018,Kirsh_2021,Frasca_2023_lin,grAl-inductors}. So far, their use has remained confined to specialized applications, with Josephson junction arrays (JJAs) continuing to serve as the more widely adopted alternatives for quantum information circuits
\,\cite{Pop2014,Nguyen_2019,Bao_2022,Somoroff_2023,Ding_2023,Lin_2025,Wang_fluxonium_2025}.

Among high-kinetic inductance materials, granular aluminum (grAl) \cite{Cohen-grAl-1968} stands out as particularly promising. Its microstructure closely resembles that of JJAs\,\cite{Deutscher1973}, inheriting many of the same appealing qualities. Furthermore, grAl fabrication is fully compatible with standard Josephson junction (JJ) fabrication techniques, easing its integration into existing device architectures\,\cite{grAl-inductors,Grunhaupt2019}. Underscoring its potential for quantum information applications, recent work has demonstrated grAl resonators with internal quality factors as high as four million\,\cite{grAl-inductors}.

\begin{figure}[t]
    \includegraphics[width = \linewidth]{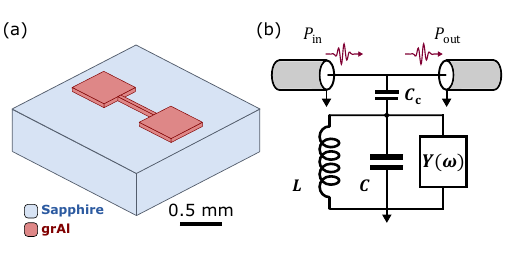}
    \caption{\textbf{Lumped-element granular aluminum microwave resonators.} (a) The sample design consists of lumped-element microwave resonators with a planar, millimeter scale two-pad capacitor geometry, connected via a central inductive strip $30\,\mathrm{\mu m}-500\,\mathrm{\mu m}$ in length\,\cite{grAl-inductors}. The devices are fabricated on sapphire substrates with superconducting granular aluminum (grAl) thin films of varying normal-state resistivity $\rho_{\mathrm{n}}$ and constant thickness $d = 91\pm1\,\mathrm{nm}$. For two devices, the capacitor electrodes are made from pure Al. (b)
    The measurement scheme of the superconducting resonator. The equivalent circuit of the resonator consists of three elements: $C$ and $L$ represent the electromagnetic contributions to the capacitance and inductance, while $Y(\omega)$ is the complex admittance that incorporates the kinetic inductance of the superconductor, the quasiparticle response, and, possibly, the contribution from TLSs. For readout, the resonators are coupled to a feedline through a capacitance $C_{\mathrm{c}}$ in a hanger-type configuration.}
    \label{fig:device}
\end{figure}

Unlike the well-documented properties of pure Al resonators\,\cite{Zmuidzinas-Resonators-Review}, 
the microwave response of grAl devices in the high-Q limit is comparatively underexplored. This gap needs to be closed to further advance the use of grAl in quantum devices. Therefore, this paper aims at a detailed investigation of the linear and nonlinear admittance $Y(\omega)$ of lumped-element grAl resonators. The ultimate goal is to identify the dominant physical mechanisms governing their behavior in the single-photon regime, and to distinguish between contributions from quasiparticles\,\cite{Charpentier2026} and defects in the surface dielectrics\,\cite{TLS-Lisenfeld-review}. To this end, we adopt a lumped-element sample design, consisting of a two-pad capacitor geometry shunted by a central inductive strip in a hanger-type readout configuration (see Fig.\,\ref{fig:device}). We study the complex admittance of grAl by characterizing the resonance frequency and loss rate as a function of temperature, $T$, and power, $P_{\text{in}}$\,\cite{Crowley_2023}.  

The microwave response of most superconducting resonators is conventionally described by the sum of two contributions: (i) Mattis-Bardeen theory \cite{Mattis-Bardeen} for the complex admittance of a uniform superconductor; and (ii) two-level system ensemble (TLS) model \cite{Gao_thesis, TLS-Gao-ApplPhysLett, TLS-Lisenfeld-review} for the dielectric permittivity of the environment. Importantly, both mechanisms lead to a decrease of the resonator frequency $f_0$ with temperature in the regime where the latter remains small, $T \ll h f_0 / k_B$. Further, one expects a negative resonance frequency shift with the increase of microwave power due to the conventional Kerr effect in a superconductor.

Here, we report the observation of an anomalous behavior in the microwave response of lumped-element grAl resonators. It manifests as an unexpected positive frequency shift with an increase in sample temperature or readout power. This anomalous behavior is observed at low temperatures (close to the base fridge temperature) and relatively low readout powers equivalent to tens of photons in the resonator. At higher $T$, the temperature dependence of the linear response parameters (frequency shift and the internal quality factor of the resonator) 
becomes conventional and is well captured by the standard Mattis–Bardeen theory, despite the granularity of the material. Overall, the frequency is a non-monotonic function of both temperature and power. In addition, the response of our devices to high-energy events, caused by cosmic ray impacts or other sources of ionizing radiation\,\cite{Day2003,IoanPop-2018,Vepsalainen2020,Gusenkova_2022,Harrington2025}, is nontrivial, exhibiting multiple relaxation timescales and a transient positive frequency shift relative to the equilibrium value.

Anomalous power or temperature dependence of the resonant frequency in various superconductors has been reported in Refs.~\onlinecite{Barends_Al,Noguchi_Al} for pure aluminum (Al), in Refs.~\onlinecite{Kristen_PhD, IoanPop-PRA-HighImpedance} for grAl, in Ref.~\onlinecite{deOry_Nb} for niobium (Nb), in Refs.~\onlinecite{Gao_2012_TiN, Chang_TiN, Swenson_2013, Gao2014_TiN} in titanium nitride (TiN), and  in Refs.~\onlinecite{Gyenis-experiment} and \onlinecite{Kirsh_2021} for tungsten silicide (WSi). In our work, we report both anomalous temperature and power dependence of the resonance frequency in the same low-loss grAl devices. 
The narrow linewidths of these devices provide the frequency sensitivity needed to resolve these effects in detail. We use multiple devices with resonance frequencies in the range of $f_0 = 5.31\text{--}7.53 \, \text{GHz}$ to thoroughly investigate this behavior; see Appendix~\ref{app:device-overview} and Tbl.~\ref{table_info} for details.

We accompany the experimental observations of the anomalies with an extensive theoretical effort to understand their origin. As mentioned above, both the Mattis-Bardeen quasiparticle theory and the phenomenological TLS theory fail to reproduce the initial positive slope of the frequency dependence on temperature.
Moreover, inclusion of the residual non-thermal quasiparticles (that remain in the sample even at the lowest temperatures) into the Mattis-Bardeen theory does not change the sign of the initial temperature dependence.
However, we show that it can be explained by the fluctuations of the superconducting gap values across the ensemble of grains comprising the material and by the nature of the electron transport associated with tunneling between the grains~\cite{Efetov-RMP, maleeva_circuit_2018}. Accounting for these peculiarities is beyond the continuous medium approximation accepted in the Mattis-Bardeen theory. 

We model granular aluminum as an array of superconducting grains with random values of the superconducting gap, connected by Josephson junctions \cite{Efetov-RMP, maleeva_circuit_2018}. The validity of this model is corroborated by the magnitude of the negative Kerr shift of frequency at higher readout powers, see Sec.~\ref{sec:evaluating-Kerr}. The observed temperature and power dependence of the resonance frequency shift can be explained by a distribution of the superconducting gap values across the array with width $\delta \Delta_0$ comparable to the resonator frequency, $\delta \Delta_0 \sim h f_0$. 

Within the same model of a network made of grains connected by Josephson junctions, our theory explains the non-monotonic dependence of resonance frequency on microwave power: increasing power heats up the quasiparticles thereby reproducing the anomalous temperature dependence. This effect saturates at larger powers, where the Kerr nonlinearity of the Josephson junctions in the network becomes dominant; the frequency decreases upon further increase of microwave power.

Despite the success of the quasiparticle theory in explaining the temperature dependence of the resonant frequency it overestimates the observed internal loss by approximately two orders of magnitude, which still calls for an explanation. We highlight a substantial disparity between the observed strengths of the temperature dependence in dissipative and non-dissipative parts of the response. The internal loss observed at $T \lesssim 200\,\text{mK}$ contributes only about $f_0/Q_\mathrm{i} \sim3$~kHz to the resonator linewidth ($Q_\mathrm{i}$ is internal quality factor), whereas the positive frequency shift reaches $50$~kHz in some devices. From the perspective of practical applications, small internal loss is favorable, although it poses a major challenge for any attempt at a theoretical explanation. 

\section{Anomalous frequency shift with temperature in granular aluminum resonators}
\label{sec:temperature-dependence}

\begin{figure}[t]
    \includegraphics[width = 0.9\linewidth]{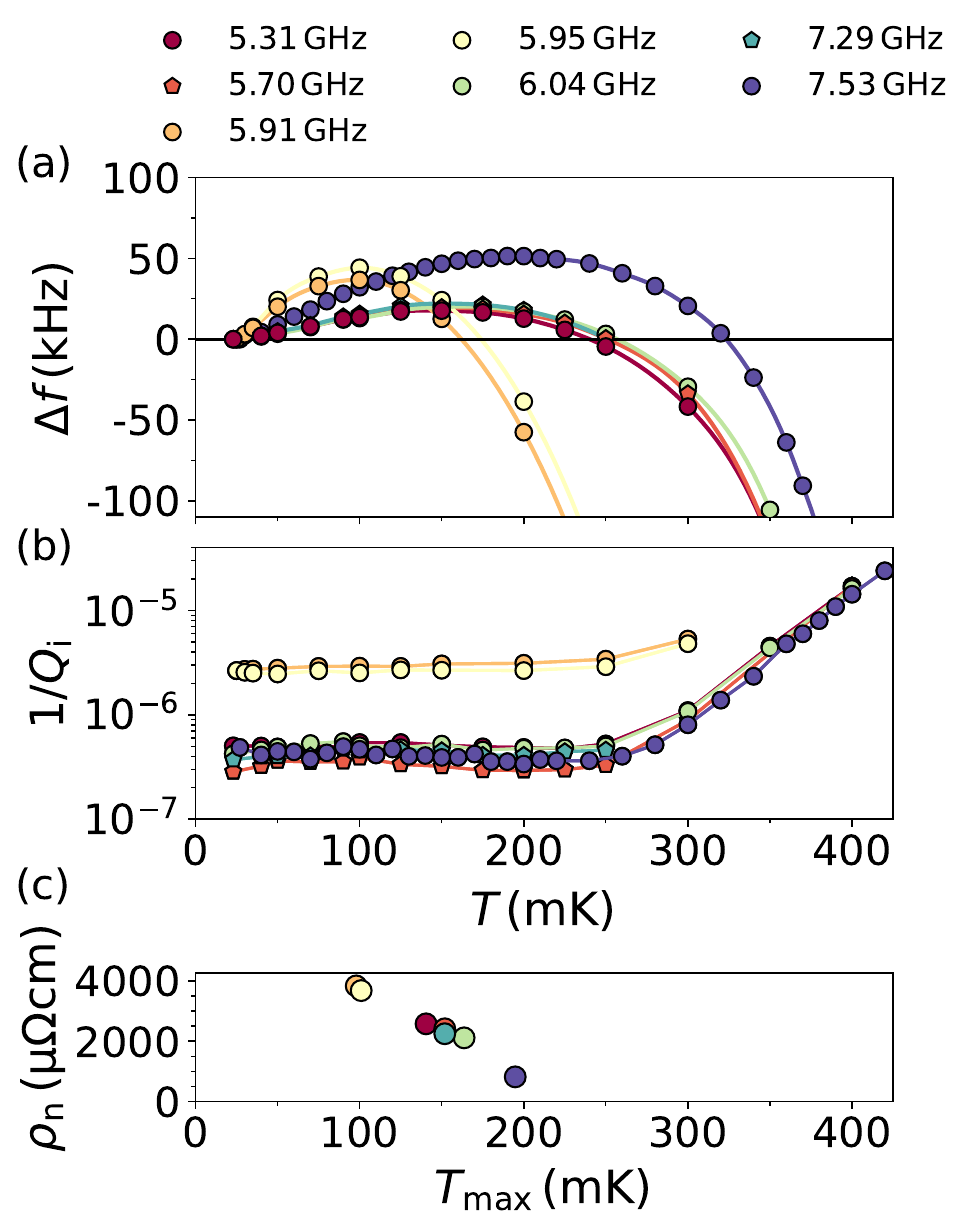}
    \caption{\textbf{Low-power temperature dependence of response at microwave frequencies:} (a) Shift in the resonance frequency $\Delta f$ as a function of the fridge base temperature $T$ measured in the single-photon regime ($\bar{n}\approx 1$) for seven devices. All curves are shifted to align them at the lowest temperature measured. In all devices, we observe the same systematic behavior: The frequency increases with increasing temperature until it reaches a maximum at $T_{\mathrm{max}}$, after which it monotonically decreases due to the thermal activation of quasiparticles (QPs). (b)  Internal loss $Q_{\rm i}$ for the same devices measured in the same temperature range. Despite the significant changes in the resonance frequency, the internal loss is independent of temperature until the thermal activation of QPs. Solid lines in panels (a) and (b) are guides to the eye. (c) Temperature $T_{\mathrm{max}}$ at which the maximal frequency shift is observed increases with the decrease of the normal-state resistivity $\rho_{\mathrm{n}}$ of the grAl film in the range of $\rho_{\mathrm{n}}$ between $500\,\mu\Omega{\rm cm}$ and $4000\,\mu\Omega{\rm cm}$. This behavior of $T_{\mathrm{max}}$ parallels that of grAl films $T_c$, which decreases 
by $\sim10\%$ in the same range of resistivity \cite{LevyBertrand-Aluminum-dome, Scheffler-Al-Dome}.  }

    \label{Fig_temperature_dependence}
\end{figure}

The temperature dependence of the resonance frequency and the internal loss are shown in Fig.\,\ref{Fig_temperature_dependence} for seven different devices measured in the single photon regime $\bar{n} \approx 1$. For a better comparison across the devices, we plot the frequency difference with respect to the lowest temperature ($T_{\min}$) value $\Delta f (T) = f_0(T) - f_0(T_{\min})$, where the approximate value of $f_0$ in each device is indicated in the corresponding legend and $T_{\min}=23\text{--}27\text{ mK}$, depending on the device and respective cool-down. 

The frequency shifts are in the range of tens to hundreds of kHz, which is much smaller than the baseline resonant frequency value of several GHz. Interestingly, all devices show an initial increase in resonance frequency at low temperatures, followed by a maximum between $100~\text{mK}$ and $200~\text{mK}$, after which the frequency decreases with increasing temperature. The internal loss rate, characterized by the inverse internal quality factor,  $1/Q_i$, seems to remain approximately constant in the same temperature range, until the thermal excitation of quasiparticles starts to dominate at temperatures above $250\,\mathrm{mK}$. 

At $T \gtrsim 350\,\text{mK}$, the temperature dependence of both the resonant frequency and the internal loss are qualitatively consistent with the contribution of thermally excited quasiparticles to the real and imaginary parts of the superconducting admittance, as described by conventional Mattis–Bardeen theory \cite{Mattis-Bardeen} with the critical temperatures $T_{\mathrm{c}}$ in the range $1.7 - 2.4\,\mathrm{K}$; see Section~\ref{sec:Mattis-Bardeen} for further details. At the same time, the nonmonotonic behavior of the resonant frequency at lower temperatures is highly nontrivial and calls for an explanation. One notable feature is the anticorrelation between the temperature at which the resonance frequency reaches its maximum (the turn-around temperature) and the normal-state resistivity $\rho_{\mathrm{n}}$ of grAl films at room temperature; see Fig.~\ref{Fig_temperature_dependence}(c). In contrast, the overall magnitude of the positive frequency shift at its maximum does not show any obvious correlation with film parameters or device geometry.

In the following Section~\ref{sec:interpretation}, we discuss in detail possible theoretical explanations for the observed behavior. In particular, we consider a commonly used TLS model for the dielectric environment and show that it is inadequate for describing the low-temperature part of the frequency dependence, as it predicts a response with both the wrong sign and wrong amplitude. We further study the effect of resident quasiparticles using a model of the granular medium. We show how that mechanism can explain the positive frequency shift and its subsequent downturn at higher temperatures. We should mention, however, that our model falls short of explaining the observed low internal loss. We therefore conclude that the anomalous temperature dependence observed here still calls for a complete theoretical explanation.

An anomalous temperature dependence has been observed previously in grAl in Ref.\,\onlinecite{Kristen_PhD}. In contrast to Ref.\,\onlinecite{Kristen_PhD}, we also report on the anomalous power dependence, which is discussed in the following Section\,\ref{sec:power-dependence}. While a positive frequency shift with readout power has been previously reported in Ref.\,\onlinecite{IoanPop-PRA-HighImpedance}, the anomalous behavior is particularly well visible in our devices thanks to their small linewidth, which highlights the unexplained difference between the change in the resonance frequency and the change in the internal loss rate.

\section{Power dependence}
\label{sec:power-dependence}

\begin{figure*}[t]
    \includegraphics[width =0.7\linewidth]{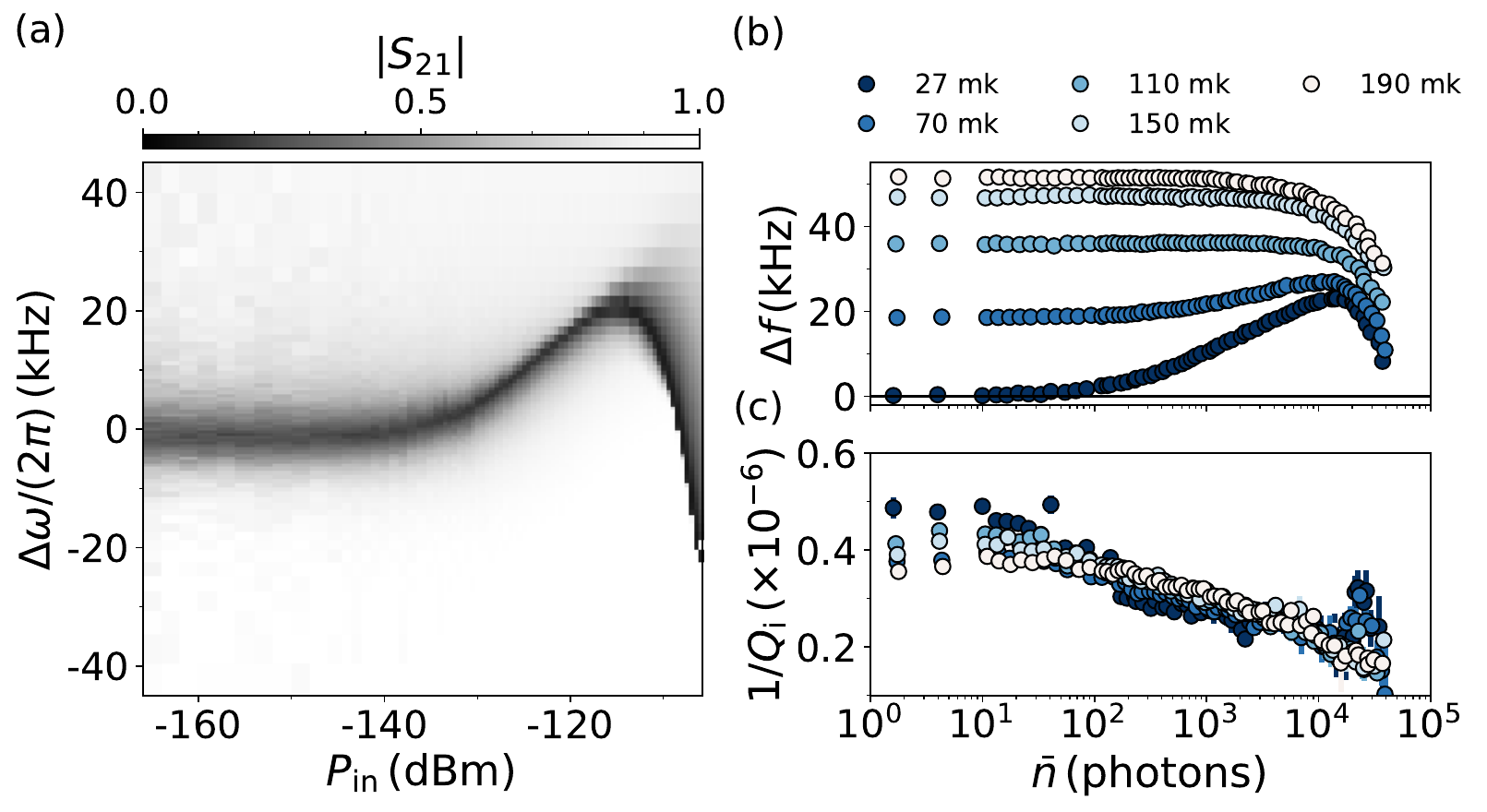}
    \caption{
    \textbf{Power and temperature dependence of the microwave response:} (a) Magnitude of the complex transmission coefficient $|S_{21}|$ at $T = 27\text{ mK}$ in the $f_0 = 7.53\text{ GHz}$ device as a function of the detuning $\Delta \omega/ (2 \pi) = f - f_0$ of the drive frequency $f$ with respect to the low power resonance frequency $f_0$ and the estimated on-chip input power $P_{\mathrm{in}}$. With increasing power, the resonance first shifts towards higher frequencies before starting to decrease, as expected. (b,c) Shift in the extracted resonance frequency $\Delta f = f_0(T,\bar{n}) - f_0(T = 27\,\mathrm{mK},\bar{n}\approx 1)$ (top panel) and internal loss rate $1/Q_{\mathrm{i}}$ (bottom panel) as a function of the mean photon number $\bar{n}$, evaluated with  Eq.~\eqref{nbar-via-P}. Different curves correspond to different values of the base temperature in the range $27\,\mathrm{mK} \leq T \leq 200\,\mathrm{mK}$. The resonance frequency shows a distinct positive shift with increasing power at low temperatures, which disappears when the temperature is raised to around $200\,\mathrm{mK}$. At the same temperatures, the low power resonance frequency reaches its maximum; see Fig.\,\ref{Fig_temperature_dependence}. As can be seen in panel (c), the power dependence of the internal loss does not change significantly with temperature. \label{Fig_f0_Qi_temp_power}
    }
\end{figure*}

In the previous section, we have introduced the anomaly observed in the temperature dependence of the resonance frequency in multiple grAl resonators. In this section, we discuss how the power dependence is also anomalous in our devices at low temperatures, and how it changes as we increase the temperature. We show a detailed study for only one of our devices with resonance frequency $f_0 = 7.53\text{ GHz}$, but a qualitatively similar behavior was observed in all samples. 

Figure~\ref{Fig_f0_Qi_temp_power} (a) shows the magnitude of the complex transmission coefficient $|S_{21}|$ as a function of the detuning $\Delta \omega = \omega - \omega_0$ between the readout tone and the resonance frequency, as well as the readout power $P_{\mathrm{in}}$, measured for the lowest device temperature. With increasing readout power, the resonance first shifts towards higher frequencies (positive detuning) up to a maximum frequency shift, above which the resonance continues to shift towards lower frequencies (negative detuning).

The amplitude of electromagnetic field in the superconducting resonator is conveniently described by the average number of photons in the resonator $\bar{n}$. In the absence of detuning, it can be evaluated through the incident power $P_{\mathrm{in}}$ as
\be \label{nbar-via-P}
    \bar{n} = \frac{2 Q_{\mathrm{e}}^{-1}}{(Q_{\mathrm{e}}^{-1} + Q_{\mathrm{i}}^{-1})^2} \frac{P_{\mathrm{in}}}{2 \pi h f_0^2},
\ee
where $Q_{{\mathrm{i(e)}}}$ are internal (external) quality factors, which are obtained from the frequency dependence of the transmission coefficient shown in Fig.~\ref{Fig_f0_Qi_temp_power} (a), as explained in the Supplementary to Ref.~\onlinecite{RMP-Quantum-measurement}. At higher powers, the Lorentzian shape of the resonance is distorted in our devices and we use the fitting to the Duffing resonator response as explained in the Appendices~\ref{Sec_Duffing} and \ref{Sec_extracting}; see Fig.~\ref{Fig_scattering_coefficient} therein. The internal and external quality factors are on the order of $10^6$ for all devices.

Figure~\ref{Fig_f0_Qi_temp_power} (b) displays the frequency shift in the $f_0=7.53\text{ GHz}$ device compared to the low-power and low-temperature reference value $\Delta f (T, \bar{n}) = f_0(T, \bar{n}) - f_0(T=27\text{ mK}, \bar{n} \approx 1)$, together with the internal loss rate. For the lowest temperature $T = 27\,\mathrm{mK}$, the resonance frequency increases as we increase the intra-cavity photon number $\bar{n}$ until it reaches a maximum after which it continues to monotonically decrease. 

The decrease in the resonant frequency observed at larger photon numbers $\bar{n}$ can be explained by the conventional Kerr nonlinearity originating from the nonlinear current-phase relation of the Josephson junctions linking the Al grains in the grAl material.  For the $f_0 = 7.53\,\text{GHz}$ device, the observed Kerr nonlinearity is $-0.54\,\text{Hz/photon}$ which is in agreement with our theoretically predicted range of $0.5\text{--}5\,\text{Hz/photon}$; see Section~\ref{sec:evaluating-Kerr} for details. The positive frequency shift at low $\bar{n}$, on the other hand, is anomalous and cannot be explained by conventional theory. Any viable explanation must involve a mechanism that overcomes the Kerr nonlinearity at low $\bar{n}$. In Section~\ref{sec:heat-balance-model}, we attempt to explain the initial part of the power dependence by attributing it to quasiparticle heating by the incident microwave drive, thereby mapping the effect onto the temperature dependence studied in Section~\ref{sec:temperature-dependence}. The results of our heat balance model can explain the  frequency dependence at powers below $-90\,\text{dBm}$. 
We mention in passing that the anomalous power dependence of the frequency shift cannot be explained by the TLS mechanism; see Eq.~\eqref{TLS-Gamma-n} and its discussion in Sec.~\ref{sec:TLSfits}.

As the base temperature of the fridge is increased, we observe an increase in the low-power frequency consistent with the observation in Fig.~\ref{Fig_temperature_dependence}(a). Additionally, the positive frequency shift with readout power is reduced with increasing temperature [see Fig.~\ref{Fig_f0_Qi_temp_power}(b)]. Note that at the same temperature at which the low-power frequency shift reaches its maximum value for this device ($T_{\max} \sim 200 \text{ mK}$, see Fig.~\ref{Fig_temperature_dependence}), the positive frequency shift with power completely vanishes too, see Fig.~\ref{Fig_f0_Qi_temp_power}(b).

For larger photon numbers $\bar{n} \sim 4 \cdot 10^4$, the resonator bifurcates, which makes it more difficult to reliably extract the internal loss rate. See Appendices\,\ref{Sec_Duffing} and \ref{Sec_extracting} for further details about this regime.

\section{High-energy events}

\begin{figure}[t]
    \includegraphics[width =1\linewidth]{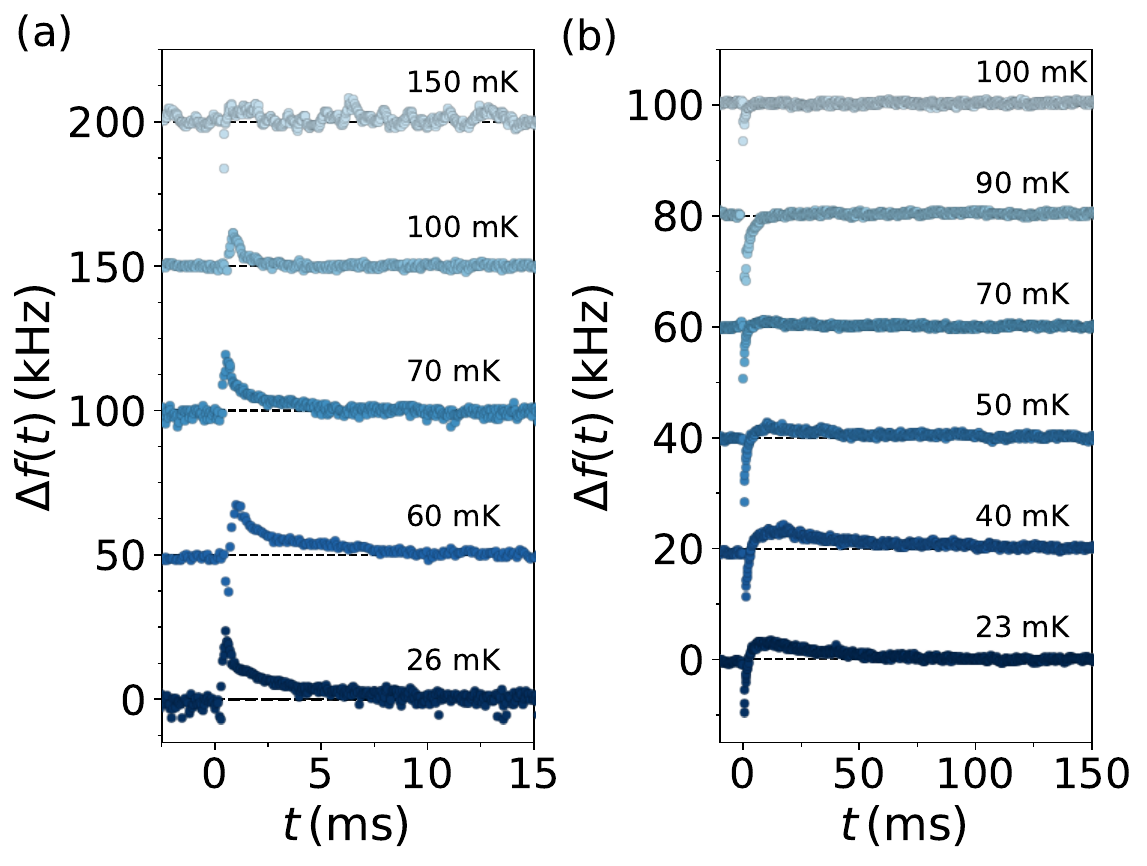}
    \caption{
    \textbf{High energy events.} Characteristic time and temperature dependence of high energy events detected in samples 7.53 GHz and 5.31 GHz. For better visibility, the curves are offset with respect to each other. Interestingly, unlike in most other superconducting resonators, the initial frequency shift can be either positive (panel a) or negative (panel b), depending on the sample. For all measured resonators, this initial spike is followed by a period of positive frequency shift, which then relaxes to the steady state over a longer timescale. In both devices, the positive frequency shift disappears with elevated base temperature, as also observed in the temperature and power dependencies discussed in Secs.~\ref{sec:temperature-dependence} and \ref{sec:power-dependence}. 
}
    \label{Fig_impacts}
\end{figure}

Another interesting observation is the response of our devices to apparent high energy events, as shown in Fig.\,\ref{Fig_impacts} for two devices measured at various temperatures. In both devices, the relaxation processes after the high energy events reveal complex dynamics with positive and negative frequency shifts with respect to the equilibrium value. Since these impacts are ultimately linked to a temporary increase in the phonon temperature followed by variations in the quasiparticles energy distribution, the dynamics observed in our devices might give insight into the relaxation behavior of the environment at different frequencies.

In the $f_0 =7.53\,\mathrm{GHz}$ device shown in the left panel, we only observe a positive frequency shift at the lowest temperatures, followed by a non-exponential relaxation back into equilibrium, see Fig.~\ref{Fig_impacts}(a). We observe two distinct time scales. First, a fast initial response to the impact, followed by a much slower relaxation on the order of a few milliseconds. In contrast, $f_0 = 5.31\,\mathrm{GHz}$ device, shown in Fig.~\ref{Fig_impacts}(b), first shows a distinct negative frequency shift, which rapidly vanishes within a few milliseconds in favor of a positive frequency shift with a much slower relaxation time on the order of tens of milliseconds. Moreover, the relaxation rate seems to be significantly smaller than in the first device. While we can only speculate about the origin of this difference, we note that the main difference between these devices is the normal-state resistivity of the grAl film and the sapphire substrate type. In either case, there are two distinct time scales in the evolution of the frequency shift, hinting at the presence of competing mechanisms causing it. A more detailed analysis of these events can be found in Appendix\,\ref{Sec_Impacts}.

\section{Data analysis and interpretation}
\label{sec:interpretation}

In the current section, we outline the theoretical frameworks used to explain and interpret the experimental data. We begin with the conventional Mattis–Bardeen theory, which successfully fits the higher-temperature part of the measurements. We then introduce the widely used TLS ensemble model and show that it fails to account for our observations. Next, we develop a granular-medium model in which resident quasiparticles contribute to the superconductor’s admittance. Finally, we estimate the conventional Kerr-nonlinearity-induced frequency shift observed at high readout powers, and present a quasiparticle-heating theory that explains the anomalous low-power part of the measurements. In this section, we focus on the $f_0 = 7.53\,\text{GHz}$ device as a representative example; the other devices have shown similar results.
To compare theoretically predicted frequency shift $\delta f$ due to each of the aforementioned mechanisms, we subtract the $T = 27\text{ mK}$ value and compare $\delta f (T) - \delta f(T = 27\text{ mK})$  with the similarly processed experimental data, $\Delta f (T)$. Note also that we aim to explain the relative frequency shifts on the order of $\Delta f / f_0\sim 10^{-5}$, which places the problem well within the perturbative regime.

\subsection{Mattis-Bardeen theory}
\label{sec:Mattis-Bardeen}

The superconducting properties of a granular material, including its electromagnetic response, are customarily described within a Josephson-junction network model \cite{Efetov-RMP}. This approach has already been successfully applied to superconducting grAl microwave resonators~\cite{maleeva_circuit_2018}. If one dispenses with the non-uniformity of the granular array, its linear response in the microwave frequency domain can be cast in the form of the standard Mattis–Bardeen theory of superconducting admittance \cite{Mattis-Bardeen}. This framework describes the complex conductivity of a superconductor in terms of its normal-state conductivity and the superconducting gap $\Delta$. It includes quasiparticle-induced corrections that can be expressed through their density normalized by the gap, $x_\mathrm{qp} = n_\mathrm{qp} / (2 \nu_0 \Delta)$, where $\nu_0$ is the electronic density of states at the Fermi level (per spin direction). In addition, the presence of quasiparticles reduces the superconducting gap according to the self-consistency equation \cite{OwenScalapino}; to the leading order, this reduction can be written as $\delta\Delta_{\text{qp}} / \Delta = - x_\mathrm{qp}$.
After some algebra, one can express the relative frequency shift in the form:
\be \label{MB-delta-f}
    \frac{\delta f^{\text{MB}}}{f_0} = - \alpha x_{\mathrm{qp}} \left(\frac{1}{2} + \sqrt{\frac{\Delta}{2 \pi k_\mathrm{B}  T}} e^{-\frac{h f_0}{2k_\mathrm{B} T}} I_0 \left(\frac{h f_0}{2 k_\mathrm{B} T}\right) \right),
\ee
where $I_0$ is the modified Bessel function of the first kind and $\alpha$ is the kinetic inductance fraction, which is assumed to be equal to one ($\alpha = 1$) in our devices~\cite{grAl-inductors}. At the same time, the quasiparticle contribution to the internal loss reads:
\be \label{MB-Qi}
    \frac{1}{Q_\mathrm{i}^{\text{MB}}} = \frac{4\alpha x_{\mathrm{qp}}}{\pi} \sqrt{\frac{\Delta}{2 \pi k_\mathrm{B} T}} \sinh \left( \frac{h f_0}{2 k_\mathrm{B} T} \right) K_0 \left(\frac{h f_0}{2 k_\mathrm{B} T} \right),
\ee
where $K_0$ is the modified Bessel function of the second kind. 

In the superconducting devices, the quasiparticle density contains a thermally activated part 
\be \label{x-qp-T}
    x_{\mathrm{qp}}^T = \sqrt{\frac{2 \pi k_\mathrm{B} T}{\Delta}} e^{-\frac{\Delta}{k_\mathrm{B} T}},
\ee
and residual density $x_{\mathrm{qp}}^{\text{res}}$, which empirically remains non-zero even at the lowest temperatures. In the simplest model, these contributions are additive:
\be \label{x-qp-full}
    x_{\mathrm{qp}} = x_{\mathrm{qp}}^{\text{res}} + x_{\mathrm{qp}}^T .
\ee

\begin{figure}[t]
    \includegraphics[width = 0.9\linewidth]{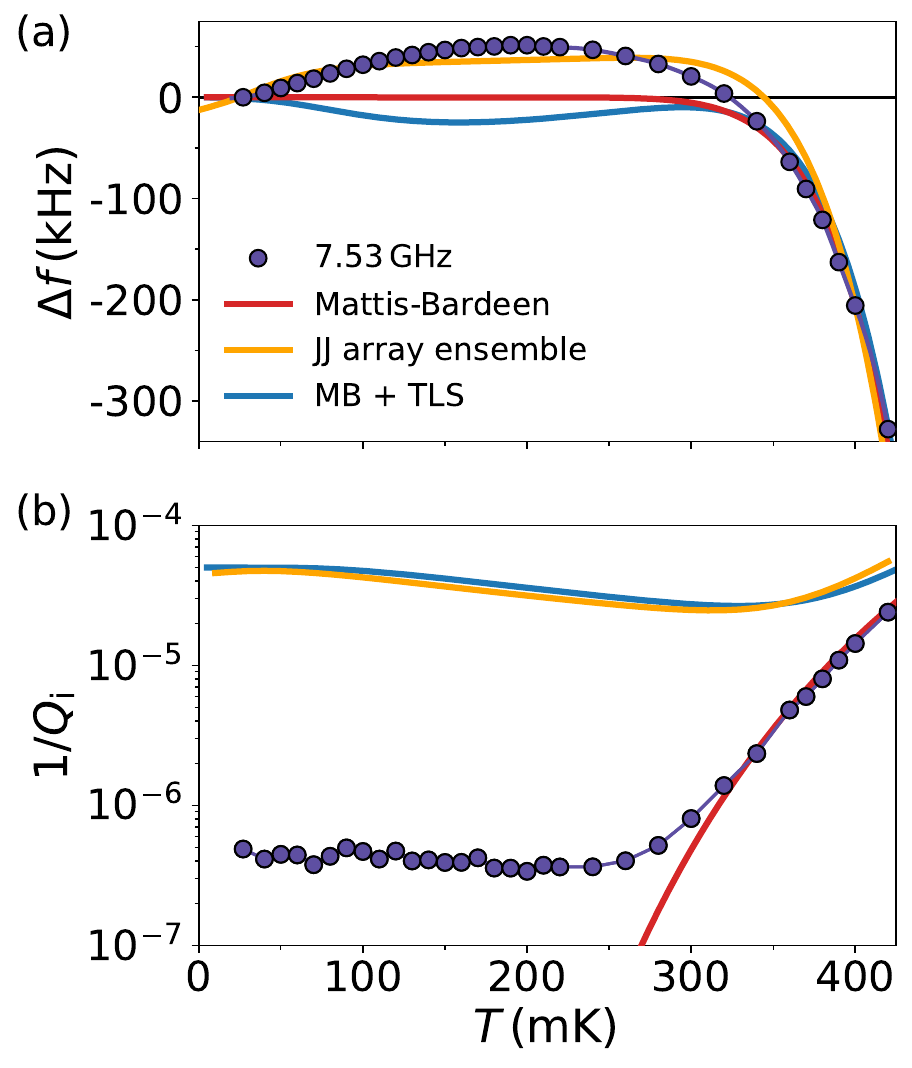}
    \caption{\textbf{Comparison of the measured temperature dependence with theoretical models.} (a) Purple dots represent the measured frequency shift $\delta f$ of the resonator, while the curves show fit attempts with various theoretical models, see the legend.  Mattis-Bardeen (MB) formula, Eq.~\eqref{MB-delta-f}, with thermally excited quasiparticles fits the data at $T \gtrsim360\,\text{mK}$ with $\Delta/h = 88 \,\text{GHz}$. Josephson junction (JJ) array model (see Sec.~\ref{sec:JJ-model}) with residual quasiparticles can decently fit the data in the whole temperature range using a random $\Delta$ described by a box distribution with the average $\Delta /h = 87\,\text{GHz}$ and width $\delta \Delta_0 / h = 1.4 f_0 = 10.5\,\text{GHz}$ at $x_{\text{qp}} = 2 \cdot 10^{-5}$. In plotting the `MB + TLS' curves, the contribution of two-level systems, Eq.~\eqref{TLS-df}, is added to the Mattis-Bardeen contribution. The prefactor $\Gamma^{\mathrm{TLS}}_{\mathrm{i,0}} = 0.5 \cdot 10^{-4}$ corresponds to the frequency shift of $\sim15\,\text{kHz}$ of a wrong sign at $T \sim 150\,\text{mK}$. (b) Contribution of the internal loss to the resonance linewidth, $f_0 / Q_\mathrm{i}$. Purple dots represent the measurements; the curves show the fits with the parameters used to fit the frequency shift [panel (a)]. Mattis-Bardeen fit matches the measurements at high temperatures and predicts zero loss at small $T$. JJ-array and TLS model fits overestimate the internal loss by two orders of magnitude at $T\lesssim100\,\text{mK}$ and by a factor of $2$--$3$ at high temperatures.
    \label{fig:theory-fits}}
\end{figure}

We find that the Mattis–Bardeen expressions, Eqs.~\eqref{MB-delta-f}–\eqref{MB-Qi}, provide a good fit to the high-temperature part of the experimental data, accurately capturing both the frequency shift and the internal loss when only thermal quasiparticles, Eq.~\eqref{x-qp-T}, are included. The fit is shown graphically in Fig.~\ref{fig:theory-fits}(a). At the same time, the anomalous low-temperature part of the frequency dependence deviates significantly from this fit, exhibiting an initial increase and a maximum before the subsequent decrease. Including a constant residual quasiparticle contribution, $x_{\mathrm{qp}}^{\text{res}}$, in the Mattis-Bardeen expressions does not reproduce the observed temperature dependence since Eq.~\eqref{MB-delta-f} is temperature-independent to the leading order at $k_\mathrm{B} T \ll \hbar \omega$, while the next-order correction produces a frequency decrease rather than an increase. Consequently, the experiment remains unexplained within the Mattis–Bardeen framework and therefore calls for an additional explanation, which we address below.

\subsection{TLS model of environment contribution}
\label{sec:TLSfits}

A two-level system (TLS) ensemble model is widely used to describe loss and dispersion in superconducting resonators and circuits \cite{Gao_thesis, TLS-Gao-ApplPhysLett, TLS-Lisenfeld-review}. TLS-like defects contribute both to the internal loss and to the resonance frequency shift and bear a temperature and readout power dependence. In this section, we review the TLS framework and show that it does not capture the low-temperature part of our observations.

A typical TLS ensemble contains two-level systems with energy splittings (frequencies) both below and above the resonator frequency $f_0$. Qualitatively, the TLS-induced shift of the resonator frequency at low temperatures can be understood in terms of level repulsion: TLS with splittings below $f_0$ ``push'' the resonator frequency up, whereas TLS with splittings above $f_0$ push it down. As the temperature increases, TLS with $h f \lesssim k_\mathrm{B} T$ become thermally saturated, and their dispersive contribution is suppressed. Therefore, a small increase in temperature first ``turns off'' the low-frequency TLS, causing the resonator frequency to decrease. Only when $k_\mathrm{B} T/h$ becomes comparable to $f_0$ do the higher-frequency TLS begin to saturate, leading to a frequency increase. Overall, coupling to a TLS ensemble produces an initial frequency drop, followed by an upturn only for $T \gtrsim h f_0/k_\mathrm{B}$, in contradiction with our observations.

Assuming a weak dependence of the TLS density of states on energy, one can derive the following analytical expression for the temperature-dependent frequency shift \cite{TLS-Hunkliger-1976, TLS-Phillips-1987, Gao_thesis, TLS-Kumar-2008}:
\begin{align} \label{TLS-df}
    \frac{\delta f^{\text{TLS}}}{f_0} = \frac{\Gamma^{\mathrm{TLS}}_{\mathrm{i,0}}}{\pi} & \text{Re} \left[\Psi\left( \frac{1}{2} + i \frac{h f_0}{2 \pi k_{\mathrm{B}} T}\right) \right. \left. - \ln \left( \frac{h f_0}{2 \pi k_{\mathrm{B}}T}\right)\right]
\end{align}
Here, $\Psi$ is the digamma function, $\Gamma^{\mathrm{TLS}}_{\mathrm{i,0}}$ is the low-power and low-temperature dissipation rate caused by the same TLS environment. In accordance with the qualitative arguments explained above, Eq.~\eqref{TLS-df} has a minimum at $T \approx 0.5 h f_0 / k_\mathrm{B}$, while the initial contribution to the frequency of the resonator is negative. 

Coupling to the TLS ensemble also contributes to the internal loss. In a phenomenological model accounting for the saturation of the transitions within a TLS by microwave photons, the dependence of the quality factor on the number $\bar{n}$ of photons in the resonator is given by
\begin{equation} \label{TLS-Gamma-n}
    \frac{1}{Q_\mathrm{i}^{\text{TLS}} (T,\bar{n})} = \Gamma^{\mathrm{TLS}}_{\mathrm{i,0}} \frac{\tanh\left( \frac{h f_0}{2 k_{\mathrm{B}}T}\right)}{\sqrt{1 +\frac{\bar{n}}{\bar{n}_{\mathrm{c}}} \tanh \left(\frac{h f_0}{2 k_{\mathrm{B}}T}\right)}}.
\end{equation}
Here $\bar{n}_c$ is the average number of photons in the resonator required for an appreciable saturation effect. As one can see, the internal loss decreases with the increase of temperature and power, because the TLS systems at the energy $h f_0$ excite and saturate to an equal population at larger temperatures or drives. Note that $\bar{n}$ did not enter Eq.~\eqref{TLS-df} since increasing power changes the population of the TLS system around the energy $h f_0$ symmetrically and does not result in a level repulsion shift. 

As explained above, the TLS ensemble fails to reproduce the correct sign of the frequency dependence at low $T$. Moreover, having a TLS-induced frequency shift with an amplitude comparable to the measured shift of $\sim50\,\text{kHz}$ will inevitably lead to a similar or greater linewidth, which contradicts the observed value of $3\,\text{kHz}$.  To illustrate its inadequacy, we plot the contribution of TLS model, Eqs.~\eqref{TLS-df} and \eqref{TLS-Gamma-n} on top of the Mattis-Bardeen fit in Fig.~\ref{fig:theory-fits} with 
value of $\Gamma^{\mathrm{TLS}}_{\mathrm{i,0}} = 0.5 \cdot 10^{-4}$, corresponding to the frequency shift of $\sim15\,\text{kHz}$, which is comparable to the observed value of the  anomalous shift but of the wrong sign.
To conclude, the TLS model cannot explain the observed temperature dependence of the frequency and quality factor in our granular aluminum resonators.

\begin{figure}[t]
    \includegraphics[width = 0.95 \linewidth]{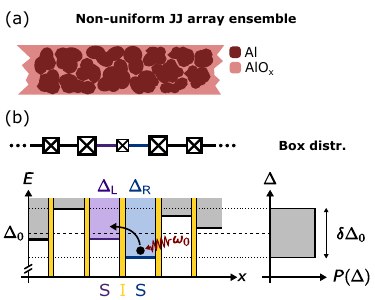}
    \caption{\textbf{Non-uniform Josephson junction ensemble model.} (a) The microstructure of superconducting grAl is composed of crystalline Al grains separated by insulating, nonstoichiometric AlOx. (b) We model the central strip of grAl in our devices with an effective ensemble of JJs with non-uniform gap parameter. Since the superconducting gap of an Al film depends on the film thickness in the range of a few tens of nanometers, we expect the gap width to vary with the grain size as well. We modeled this variation of the gap in the grains by sampling the gap in the islands of the JJ array from a uniform box distribution of width $\delta \Delta_0$ around the mean value $\Delta$.
    }  \label{Fig_JJ_ensemble_model}
\end{figure}

\subsection{Quasiparticle effect in granular media with $\Delta$ nonuniform across the array of grains}
\label{sec:JJ-model}

In this Section, we study the frequency shift and dissipation due to quasiparticles, which are ubiquitously present in superconducting devices. We are able to explain the positive frequency shift within a model of granular media accounting for a distribution of the gap widths across the ensemble of grains comprising the material. The anomalously small internal loss, however, remains unexplained.

As mentioned above, in real devices the quasiparticle density at 
low temperatures saturates at a certain device-dependent ``resident quasiparticles'' level, which is far above the equilibrium value at the same temperature (and zero chemical potential). We assume the resident quasiparticle density  $x_{\mathrm{qp}}^{\text{res}}$ is independent of temperature and contributes additively to the total quasiparticle density, see Eq.~\eqref{x-qp-full}. We saw in Sec.~\ref{sec:Mattis-Bardeen} that Mattis-Bardeen theory for a superconductor with a uniform gap $\Delta$ predicts the negative sign for the temperature dependence of the resonant frequency, even if one includes the presence of resident quasiparticles. Therefore, we account now for the spatial fluctuations of $\Delta$ across the granular material. Moreover, since we are describing a granular aluminum film, we use the Josephson junction array model. Since grains vary in size, we assume that $\Delta$ in each grain is different. We model this variation by a ``box'' distribution with an average $\Delta$ and width $\delta\Delta_0$. 

As a starting point, we consider a single asymmetric Josephson junction between two grains with superconducting gaps $\Delta_\mathrm{L}$ and $\Delta_\mathrm{R}$, differing from each other ({\sl e.g.},  $\Delta_\mathrm{R}>\Delta_\mathrm{L}$). Such a junction, with a quasiparticle capable of tunneling between the leads, acts as a somewhat unusual TLS. A quasiparticle initially residing at energy $\varepsilon$ near the edge of the continuum in the left  ($L$) lead, can be promoted by absorption of a photon to a higher-energy state in the right lead at any $\varepsilon>\Delta_\mathrm{R}-\Delta_\mathrm{L}- hf_0$ (here we measure $\varepsilon$ from the bottom of the respective quasiparticle continuum). Moreover, at $\varepsilon-(\Delta_\mathrm{R}-\Delta_\mathrm{L}- hf_0)\to +0$ the quasiparticle transition amplitude is enhanced by the singularity in the density of states near the edge of the continuum in lead  $R$, while at $\varepsilon-(\Delta_\mathrm{R}-\Delta_\mathrm{L}- hf_0)\to -0$ the amplitude is zero as there are no available final states for the transition. This asymmetry (with respect to frequency $f_0$ at fixed other parameters) results in an asymmetric ``push" of the resonator frequency. The frequency is pushed down if $hf_0<\Delta_\mathrm{R}-\Delta_\mathrm{L}-\varepsilon$, and there is a suppressed effect on the frequency in the opposite case. Now we see the emergence of $T_*=\Delta_\mathrm{R}-\Delta_\mathrm{L}-hf_0$ as a new temperature scale~\cite{Theory-qp-asymmetric}, which should be compared with the typical quasiparticle energy $\varepsilon\sim k_\mathrm{B} T$.
The push-down effect is strong at $T \lesssim T_*$, when the probability of a sufficiently small $\varepsilon$ is high, and gradually decreases when the temperature is raised above $0.6 T_*$.
Averaging over the box distribution of $\Delta_\mathrm{R}$ and $\Delta_\mathrm{L}$ enhances this effect leading to the positive frequency dependence on temperature starting from the smallest temperatures as long as $hf_0 < \delta \Delta$, see the yellow curve in Fig.~\ref{fig:theory-fits}(a).

\begin{figure}[t]
    \includegraphics[width = 0.85 \linewidth]{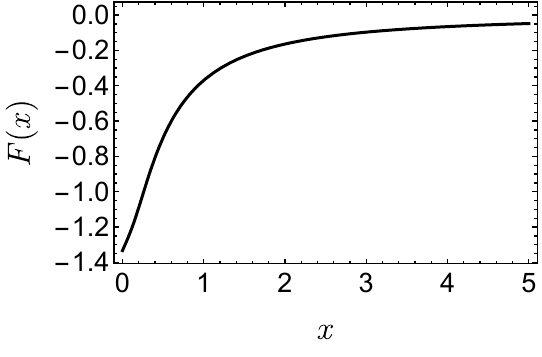}
    \caption{Function $F(x)$ [see Eq.~\eqref{ddmw-contribution}] describing the singular contribution to the frequency shift from the quasiparticles in the Josephson-junction array model of Sec.~\ref{sec:JJ-model}. Here $x$ stands for rescaled temperature: $x = k_B T / (\delta\Delta_0 - hf_0)$. This contribution leads to the positive frequency shift and dominates in the temperature dependence when $(\delta\Delta_0 - hf_0) / (hf_0) \ll 1$.
}
    \label{fig:F-x-function}
\end{figure}

The quasiparticle contribution varies between different junctions but remains small compared to the main part of the superconducting response provided that $x_{\mathrm{qp}} \ll 1$. This smallness simplifies the procedure of averaging over the random distribution of the superconducting gap widths of the individual grains. Further details of this procedure can be found in Ref.~\onlinecite{Theory-qp-random}, where we demonstrate that the resulting expression leads to a positive frequency shift at sufficiently small and positive values of the ratio $D =(\delta\Delta_0 - hf_0) / (hf_0)$. In the limit $D \ll 1$, the resulting contribution to the frequency shift can be written in the form:
\be \label{ddmw-contribution}
	\frac{\delta f^{\text{JJ-qp}}}{f_0} =  \frac{3 x_{\mathrm{qp}}}{4\sqrt{2}\pi} \frac{\sqrt{\Delta(\delta\Delta_0 - 
h f_0)}}{\delta\Delta_0} F \left( \frac{k_\mathrm{B} T}{\delta \Delta_0 - h f_0} \right),
\ee
where the function $F(x)$ can be expressed through the modified Bessel function of the first kind: ${F(x) = -(8\sqrt{\pi} / 3) x^{3/2} e^{-x/2} I_1(x/2)}$; we show its plot in Fig.~\ref{fig:F-x-function}. It has asymptotes $F(x) \sim -4/3 + x$ at $x \ll 1$ and $F(x) \to 0$ at $x \gg 1$. 

For the purposes of the present work, we use the full expression for $\delta f^{\text{JJ-qp}}$ at arbitrary $D$, which generalizes \eqref{ddmw-contribution}, see Ref.~\cite{Theory-qp-random} for the details. We include the contribution of thermal quasiparticles to $x_{qp}$, discussed in Sec.~\ref{sec:Mattis-Bardeen} and the superconducting gap suppression caused by them, $\delta\Delta_T / \Delta = - x_\mathrm{qp}^T$. Using the resulting expression as a fitting formula with three parameters: $\Delta$, $\delta \Delta_0$, and $x_{\mathrm{qp}}^{\text{res}}$, we analyzed the experimental data. The results for the $f_0 = 7.53\,\text{GHz}$ resonator are presented in Fig.~\ref{fig:theory-fits}(a) and demonstrate a decent fit of the experimental data at $\Delta / h = 87\,\text{GHz}$, $\delta \Delta_0 / h = 1.4 f_0$ ($10.5\,\text{GHz}$) and $x_{\mathrm{qp}}^{\text{res}} = 2 \cdot 10^{-5}$.

Turning to the question of the internal loss, we also evaluate its quasiparticle-induced part by averaging the dissipative part of the asymmetric Josephson junction admittance \cite{Theory-qp-asymmetric} over random $\Delta$ in the superconducting granules \cite{Theory-qp-random}. Using the fitting parameters obtained above, we evaluate the internal contribution to the linewidth and plot it in  Fig.~\ref{fig:theory-fits}(b). As one can see, the theory overestimates the linewidth by two orders of magnitude at low temperatures. We conclude that quasiparticles cannot fully explain the observed measurements and additional study is necessary.

\subsection{Evaluation of the negative Kerr shift due to the nonlinearity}
\label{sec:evaluating-Kerr}
In this Section, we estimate the negative frequency shift of the resonator with increasing drive power, arising from the nonlinear response of the granular superconductor. We associate the frequency shift with the Kerr nonlinearity of the Josephson junctions between the grains comprising grAl. This effect is observed experimentally at large photon numbers $\bar{n}$, while at small $\bar{n}$ it is overshadowed by the anomalous positive frequency shift, which we address below in Sec.~\ref{sec:heat-balance-model}.

We model granular aluminum as a network of effective Josephson junctions between individual superconducting grains~\cite{Efetov-RMP, maleeva_circuit_2018}. To estimate the Kerr effect, one can consider a single (one-dimensional) chain of junctions because inclusion of all parallel chains does not change the relative value of the frequency shift. 
The phase difference across the superconductor is divided into $N_\mathrm{J}$ jumps on each individual junction (for simplicity, we consider phase jumps uniform along the chain). Here $N_\mathrm{J}$ is the typical number of junctions between the terminals and can be estimated as $N_\mathrm{J} =l_{\text{strip}} / d_\mathrm{g}$, where $l_{\text{strip}}$ is the length of grAl segment and $d_\mathrm{g} = 5$--$15\,\text{nm}$ is the size of its grain \cite{Yang-STM}. Then, the relative frequency shift can be estimated as
\be \label{Kerr-shift}
    \frac{\delta f_\mathrm{K}}{f_0} = -\sqrt{\frac{E_\mathrm{C}}{8E_\mathrm{J}}} \frac{\bar{n}}{N_\mathrm{J}^2},
\ee
where $E_\mathrm{J}$ and $E_\mathrm{C}$ are effective net Josephson and charging energies. Note that the width $w$ of the superconductor does not enter the estimate for the relative frequency shift, since it merely increases the number of Josephson junction chains connected in parallel. 

The value of the Josephson energy $E_\mathrm{J}$ can be obtained from the Ambegaokar-Baratoff \cite{AmbegaokarBaratoff} expression for the kinetic inductance,
\be
    E_J = \frac{\Delta}{8(e^2/h)R}.
\ee
Here the normal-state resistance is $R = N_{\text{sq}} \rho_n / d$, where $\rho_\mathrm{n}$ is normal-state resistivity, $d$ is the grAl thickness, and $N_{\text{sq}}$ is the $l/w$ is the ``number of squares'' in the grAl strip. For the $l = 160\,\mu\text{m}$ and $w = 2\,\mu\text{m}$ device, we estimate $R$ as $7.2\,\text{k}\Omega$ using the room temperature resistivity $\rho_n = 820\,\mu\Omega\text{cm}$ (based on the previous literature \cite{Scheffler-Al-Dome, Deutscher-OpticalConductivity,Bachar_2013}, we expect this value to be close to the one at $4 \text{ K}$).

The charging energy $E_\mathrm{C}$ can be obtained from the relation to the unperturbed frequency of the resonator, using the relation $h f_0 = \sqrt{8 E_\mathrm{J} E_\mathrm{C}}$. Alternatively, one may obtain similar estimates for $E_C$ from electromagnetic simulation of the resonator geometry, as was done in Ref.~\onlinecite{grAl-inductors} producing 
results of the same order of magnitude .
After some algebra, we estimate the Kerr shift to be in the range $0.5$--$5\,\text{Hz/photon}$, depending on the effective grain size. 
This range matches the value of $K_0/(2\pi) = - 0.54\,\text{Hz/photon}$, obtained from the measurements by fitting the large-power part of the frequency dependence, see Fig.~\ref{fig:Pdependence-fit}(a).

\subsection{Heat balance model for the low-$T$ part of the power dependence}
\label{sec:heat-balance-model}

\begin{figure}[t]
        \centering
                \includegraphics[width = 0.95\linewidth]{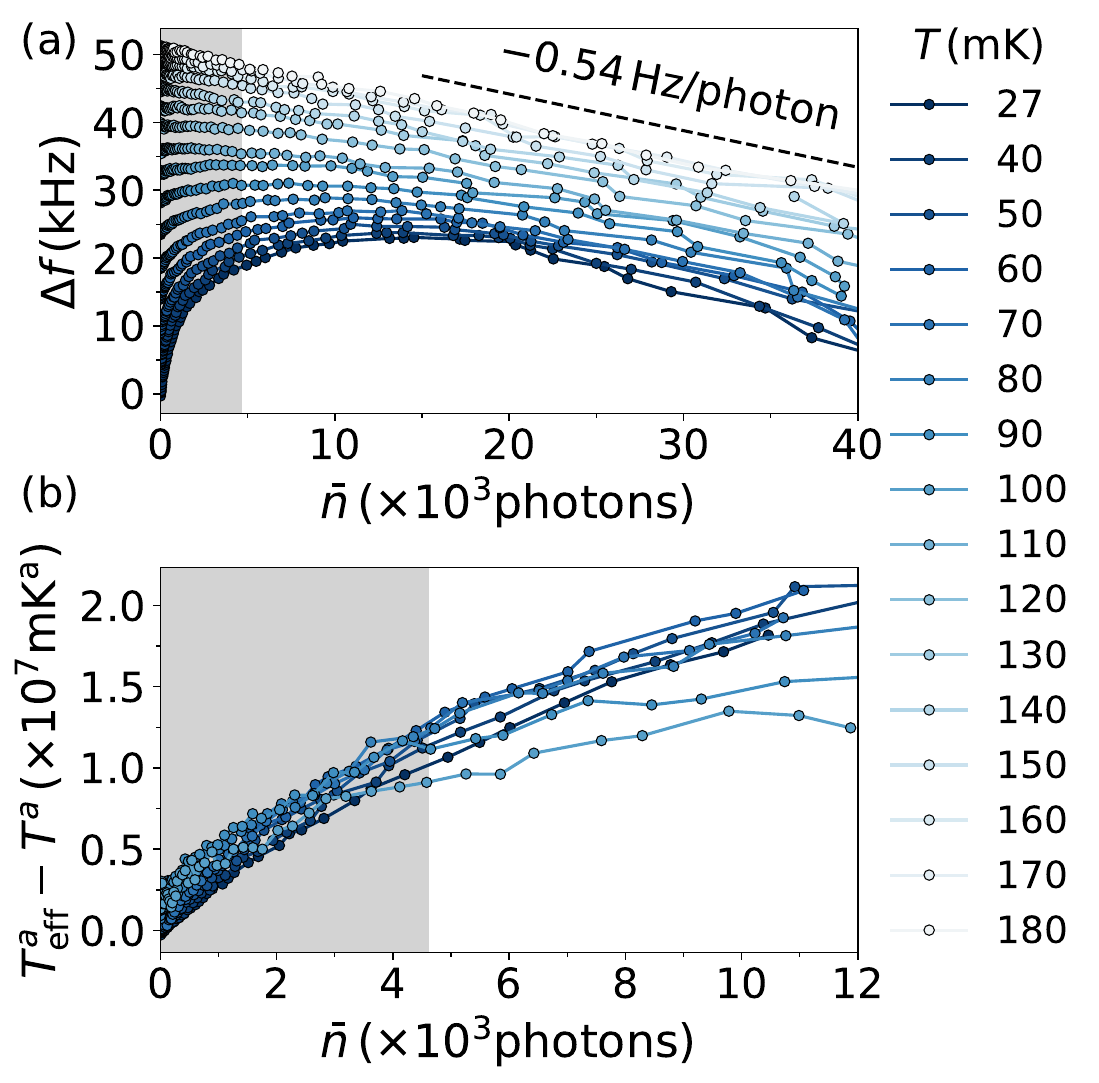}
        \caption{
        \textbf{Theoretical analysis of the dependence of the resonator frequency on power.}
        (a) Resonator frequency shift, $\Delta f $, vs. drive power, characterized by the mean intracavity photon number, $\bar{n}$, for the $f_0=7.53\text{ GHz}$ device at fridge temperatures $T=27 - 180\text{ mK}$. At larger powers, the frequency decreases linearly with a slope of $K_0 / (2\pi) = -0.54\text{ Hz/photon}$ (dashed line), which we attribute to the Kerr nonlinearity of the Josephson-junction array used to model the granular aluminum medium; see Sec.~\ref{sec:evaluating-Kerr}. The observed value of $K_0/(2\pi)$ lies within the theoretical estimate of $0.5 - 5 \text{ Hz/photon}$.
        (b) Data analysis of the power dependence in the quasiparticle heating model; see Sec.~\ref{sec:heat-balance-model}. We find the quasiparticle effective temperature $T_{\text{eff}}$ at photon number $\bar{n}$ by mapping the corrected frequency shift $\Delta f - K_0 \bar{n} / (2\pi)$ to the $\Delta f$ vs. $T$ dependence at the lowest power, see Fig.~\ref{Fig_temperature_dependence}. According to Eq.~\eqref{nbar-vs-T-theory} representing the theoretical model, all curves should collapse onto a single linear dependence in the axes of $T_{\text{eff}}^a - T^a$ vs. $\bar{n}$. Here the exponent $a$ is the only fitting parameter. The data collapse works reasonably well at $T < 100\text{ mK}$ and $\bar{n} < 4.5\cdot10^3$ (grey-shaded part of the plot) for $a=3.8$; see Sec.~\ref{sec:heat-balance-model} for interpretation. 
        \label{fig:Pdependence-fit}
        }
\end{figure}

Here we present a semi-phenomenological theory that explains the anomalous positive frequency shift with power as reported in Sec.~\ref{sec:power-dependence}; see Fig.~\ref{Fig_f0_Qi_temp_power}. It attributes this effect to heating of the quasiparticle subsystem by the applied drive. 

As we know from the measurements of the temperature dependence in the linear regime (Sec. \ref{sec:temperature-dependence}), the resonator frequency increases with increasing temperature. A part of the dissipated power of the electromagnetic field is transferred directly to the quasiparticles, 
\be \label{P-dissipative-via-n-bar}
    P_{\text{diss,qp}} =  2 \pi \beta h f_0^2 \frac{\bar{n}}{Q_\mathrm{i}},
\ee
where $\beta$ is a fraction of the internal loss due to quasiparticles. Dissipation increases the effective temperature of the quasiparticles $T_{\text{eff}}$ above the temperature $T$ set by the cryostat. The balance of power delivered to the quasiparticles and transferred from quasiparticles to other degrees of freedom determines the value of $T_{\text{eff}}$.

Assuming that the phonon mechanism is dominant and approximating the quasiparticle energy distribution by a quasi-equilibrium one with the effective temperature $T_{\text{eff}}$ and fixed value of $x_{\mathrm{qp}}=x_{\mathrm{qp}}^{\rm res}$, we write the heat balance equation in the quasiequilibrium:
\be \label{P-balance}
    P_{\text{diss,qp}} (\bar{n}) = P_{\text{e-ph}}(T_{\text{eff}}, T).
\ee
This equation implicitly determines the function of the effective quasiparticle temperature on the photon number, $T_{\text{eff}}(\bar{n})$. 

The electron-phonon power transfer $P_{\text{e-ph}}(T_{\text{eff}}, T)$ is often well-approximated \cite{Pekola-HeatFlowReview} by a power-law expression
\be \label{P-e-ph}
    P_{\text{e-ph}}(T_{\text{eff}}, T) = \Sigma \mathcal{V} \left(T_{\text{eff}}^{a} - T^{a} \right),
\ee
where $\mathcal{V}$ is the volume of the system, $\Sigma$ is a material constant depending on the electron-phonon coupling, and $a$ is a number that depends on the dimensionality, strength of disorder in the system, and its type. For a uniformly disordered quasi-two-dimensional superconducting film $a = 5.5$ \cite{SavichGlazmanKamenev}, while in certain granular superconductors a smaller value of $a \sim 3.5$ has been reported \cite{Goltsman-NbN, SkvortsovStepanov-GranularTheory}. 

Equations~\eqref{P-balance} and \eqref{P-e-ph} together determine the effective quasiparticle temperature in the system. Clearly, the effective temperature $T_{\rm eff}$ increases with $\bar n$. From low-power ($\bar n\to 0$) measurements, we know that the resonance frequency at small $T$ increases with temperature. If the only effect of a higher photon number $\bar n$ were to raise the temperature, $T\to T_{\rm eff}$, then the resonance frequency would increase with $\bar n$ (at fixed $T$) as well.

For a quantitative prediction of frequency on power, one would need to know the material parameters $\Sigma$ and $a$, as well as the coefficient $\beta$. However, one can also check the validity of the model without knowing the parameter values a priori. 

To this end, we first subtract the conventional negative frequency shift, which is observed at higher drive powers. It was addressed in Sec.~\ref{sec:evaluating-Kerr} and interpreted as a Kerr effect in the granular medium with a linear dependence on the photon number, $\delta f_\text{K} = K_0 \bar{n}/(2\pi)$. We extract the anomalous part of the shift, $\widetilde{\Delta f}$, by subtracting $\delta f_\text{K}$ from the experimental data for the frequency shift $\Delta f$, obtained as a function of $\bar n$ and $T$ from the high-power measurements, see Fig.~\ref{fig:Pdependence-fit}(a). We then attribute $\widetilde{\Delta f} = \Delta f - K_0 \bar{n} / (2\pi)$ to the increase of effective temperature $T_{\rm eff}$ with $\bar n$. The identification of 
$\widetilde{\Delta f}$ with the independently measured $\Delta f$ vs. $T$ dependence in the linear regime ($\bar{n}\ll 1$), allows one to find the effective temperature $T_{\text{eff}}(T, \bar{n})$. If the assumptions of the theory are right, then the entire set of $T_{\text{eff}}(T, \bar{n})$ data should satisfy Eqs.~\eqref{P-dissipative-via-n-bar}--\eqref{P-e-ph}, which can be abbreviated to
\be \label{nbar-vs-T-theory}
    \bar{n} = C [ T_{\text{eff}}^a - T^a ].
\ee 
The constant $C = Q_\mathrm{i} \Sigma \mathcal{V} / (2\pi \beta h f_0^2)$ here depends on the material parameters, but---crucially---it is independent of $T$ or $\bar n$. Therefore, with a proper $a$, all data for power sweeps at each cryostat temperature $T$ should collapse onto a single straight line when plotted as $T_{\text{eff}}^a - T^a$ vs. $\bar{n}$. As we can see from Fig.~\ref{fig:Pdependence-fit}(b), that holds reasonably well for the powers $\bar{n} \lesssim 4.5 \cdot 10^3$ and fridge temperatures $T < 100 \text{ mK}$ with the fitted value $a = 3.8$.

\section{Discussion}
In summary, we have reported an anomalous positive frequency shift in low-loss lumped element resonators made from grAl with increasing sample temperature or microwave power. This behavior is observed in the low-temperature domain and cannot be explained within the conventional framework that combines the Mattis-Bardeen theory with the phenomenological TLS theory to account for the conductor and dielectric responses, respectively. Similarly, the anomalous power dependence in the low-power domain cannot be explained by the Kerr nonlinearity of Josephson junctions connecting Al grains within the grAl material. We are able, however, to explain these anomalies by accounting for the randomness of the superconducting gaps in the ensemble of grains comprising grAl. 

In addition to accounting for the low-temperature anomaly, our theory for the temperature dependence of the resonance frequency smoothly connects with the conventional Mattis-Bardeen theory in the limit of large temperature. However, the largest measured temperatures fall into the crossover region, where our theory overestimates the internal loss by a factor of $2\text{--}3$, see Figure~\ref{fig:theory-fits}. In that higher-temperature interval, conventional Mattis-Bardeen theory provides a better fit as it adequately describes both the measured non-dissipative and dissipative parts of the microwave response. 

Interestingly, the observed anomalous low-temperature behavior of the resonance frequency is not accompanied by a commensurate change in the internal loss rate.
A positive frequency shift on the order of $50\,\text{kHz}$, combined with an internal contribution to the linewidth of only $\sim3\,\text{kHz}$, poses a serious challenge for any theory aiming to explain the totality of experimental data. Any dynamical response of the material or its environment, whether quasiparticle-related or dielectric, is constrained by the Kramers–Kronig relations~\cite{Kronig_1926,Kramers_1927}, which imply that once the temperature dependence becomes sizable, the dissipative component must be at least comparable to the reactive one. A tempting way to resolve this apparent paradox is to invoke a purely inductive contribution to the admittance, $Y_{\text{ind}} \sim -i/\omega$, which is not directly constrained by the Kramers–Kronig relations. However, in the relevant models discussed in Sec.~\ref{sec:interpretation}, we find that the major part of the inductive contribution is temperature-independent. Consequently, fitting the measured frequency shift with the full theoretical expression forces the model into a regime where dissipative processes become significant. That leads to a predicted dissipation that greatly exceeds the experimentally observed values. Resolving this problem is an open question that we intend to address in our future work.

\begin{acknowledgements}

We thank Ioan M. Pop, Thomas Reisinger, and Harvey Moseley for fruitful discussions. This research was sponsored by the Army Research Office (ARO) under grant no. W911NF-23-1-0051, and by the
U.S. Department of Energy (DoE) grant no. DE-SC0012704, Office of Science, National Quantum Information Science Research Centers, Co-design Center for Quantum Advantage (C2QA).
The views and conclusions contained in this document are those of the authors and should not be interpreted as representing the official policies, either expressed or implied, of the ARO, DoE, or the US Government. The US Government is authorized to reproduce and distribute reprints for Government purposes notwithstanding any copyright notation herein. Fabrication facilities use was partially supported by the Yale University Cleanroom, a core facility under the directorate of the Provost Office. We thank Yong Sun, Lauren McCabe and Kelly Woods for guidance and assistance towards developing and implementing fabrication processes. We also acknowledge the support of the Yale Quantum Institute. 

L.F. and R.J.S. are consultants and shareholders of D-Wave Quantum.
\end{acknowledgements}


\appendix

\vspace{1cm}

\begin{table*}[t!]
\centering
\caption{\textbf{Summary of resonator parameters, measured frequencies and quality factors:} The wafer label indicates the different batches used in our study, while the substrate column indicates the sapphire type. We also give the cool-down (CD) index and the device type describes the materials used. The other resonator parameters are the sheet resistance $R_{\mathrm{sq}}$ measured at room temperature, the normal-state resistivity $\rho_{\mathrm{n}} = R_{\mathrm{sq}} d$ calculated from the sheet resistance and the film thickness, as well as the length and the width of the strip $l_{\mathrm{strip}}$ and $w_{\mathrm{strip}}$, respectively. The kinetic inductance fraction $\alpha = 1 - L_{\mathrm{g}}C_{\mathrm{s}}\omega_{\mathrm{0}}^2$ is estimated from the measured resonance frequency $f_0$ and the simulated geometric inductance $L_{\mathrm{g}}$ and shunt capacitance $C_{\mathrm{s}}$. The internal quality factor $Q_{\mathrm{i}}$ is given in the single-photon regime and the lowest fridge temperature $\sim23\text{--}27\text{ mK}$. All resonators presented in this table nominally have the same grAl film thickness of $d = 91\pm1$\,nm.
} 
\begin{tabular}{|c|c|c|c|c c|c c c|c|c c|c|c|}
\hline

\textbf{Wafer} &
\textbf{Substrate} &
\textbf{CD index} &
\textbf{Type} &
\textbf{$R_{\mathrm{sq}}$} &
\textbf{$\rho_{\mathrm{n}}$} & 
\textbf{$l_{\mathrm{strip}}$} &
\textbf{ $w_{\mathrm{strip}}$} & 
\textbf{$f_0$} & 
\textbf{$\alpha$} &
\textbf{$Q_{\mathrm{i}}$ } &
\textbf{$Q_{\mathrm{e}}$ }  &
\textbf{marker}\\

 & & & & $(\Omega)$ & $(\si{\micro\Omega}$cm) & $(\si{\micro\metre})$ & $(\si{\micro\metre})$ & (GHz) & & $(\times 10^6)$ & $(\times 10^6)$ & \\
\hline 
\hline

FN23  & EFG & 5 & all-grAl & 421 & 3831 & 30 & 2 & 5.91 & $>0.95$ & 0.35 & 1.55 & \includegraphics[height = 0.25cm]{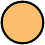} \\

FN23  & EFG & 5 & all-grAl & 404 & 3676 & 34 & 2 & 5.95 & $>0.95$ & 0.35 & 0.51 & \includegraphics[height = 0.25cm]{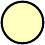} \\

\hline
\hline

FO23  & EFG & 11 &all-grAl & 232 & 2111 & 150 & 3 & 6.04 & 0.94 & 1.78 & 1.30 & \includegraphics[height = 0.25cm]{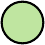} \\

\hline
\hline

I24   & EFG & 11 & all-grAl & 283 & 2575 & 500 & 10 & 5.31 & 0.92 & 1.45 & 1.08 & \includegraphics[height = 0.25cm]{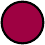} \\

I24 & EFG & 11 & hybrid: grAl/Al & 265 & 2412 & 400 & 8 & 5.70 & 0.94 & 2.5 & 0.57 & \includegraphics[height = 0.25cm]{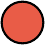} \\

I24 & EFG & 11 & hybrid: grAl/Al & 247 & 2247 & 294 & 10 & 7.29 & 0.90 & 2.11 & 1.71 & \includegraphics[height = 0.25cm]{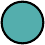} \\

\hline
\hline

BM24 & HEM & 15 & all-grAl & 90 & 819 & 160 & 2 & 7.53 & 0.90 &  1.85 & 0.6 & 
\includegraphics[height = 0.25cm]{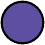} \\

\hline
\end{tabular}  
\label{table_info}
\end{table*}

\section{Device overview}
\label{app:device-overview}
The device and grAl film parameters are summarized in Tbl.\,\ref{table_info} below. All devices are fabricated either on an Edge-fed Film Growth (EFG) or Heat Exchanger Method (HEM) sapphire substrate and patterned using photolithography. Four out of the seven samples (batches FO23 and I24) were already reported in Ref.\,\cite{grAl-inductors}, with a detailed description of the fabrication recipe and measurement setup in Appendices B and D of the same publication, respectively. The only noteworthy difference in the fabrication of batch FN23 is the use of an additional in-situ ``cleaning'' step before the grAl film deposition. 
We stopped using this cleaning step since it involved Ar milling, which was potentially increasing the substrate surface roughness. The thickness of the grAl films was measured with a Tecor P-7 Stylus Profiler.

The data was collected in three separate cool downs (CD) with internal reference index 5, 11, and 15. Five out of the seven samples are all-grAl devices, for which the entire circuit is fabricated from a single layer of grAl, while two samples are hybrid samples for which the capacitor electrodes are made from pure Al. While we have observed an anomalous power dependence in all devices reported in Ref.\,\cite{grAl-inductors}, we only discuss a sub-set here for which we have also measured the temperature dependence.

We calculate the kinetic inductance fraction $\alpha = 1 - L_{\mathrm{g}}C_{\mathrm{s}}\omega_{\mathrm{0}}^2$ from the measured resonance frequencies $\omega_0$, and the simulated geometric inductance and shunt capacitance $L_{\mathrm{g}}$ and $C_{\mathrm{s}}$, respectively. For our simulations, we use the 3D High Frequency Structure Simulation (HFSS) software from Ansys.  

\section{Transmission coefficient for the Duffing oscillator}
\label{Sec_Duffing}
In this section, we derive an analytical expression for the transmission coefficient of the Duffing oscillator as a function of the readout frequency and amplitude, which we use to extract the resonance frequency together with the internal and external loss rates.

The Duffing oscillator is an extension to the simple harmonic oscillator, and describes a driven, damped oscillator with non-harmonic, quartic potential\,\cite{Duffing_1921, Nonlinear_oscillations_book}. Expressed in the language of quantum optics, the Hamiltonian of the Duffing oscillator is\,\cite{Imoto_1985,Yurke2006}
\begin{equation} \label{Duffing-Hamiltonian}
    H/\hbar = \omega_0 a^\dagger a + \frac{K}{2} a^\dagger a^\dagger a a.
\end{equation}
Here, $a^\dagger$ and $a$ are the single-mode field amplitude creation and annihilation operators, $\omega_0$ is the resonance frequency and $K$ is the Kerr coefficient. The energy difference between neighboring energy levels decreases by $K$.

\subsection{Quantum Langevin equation of motion in the hanger geometry}
In this section, we derive an expression for the steady-state solution of the intra-cavity field of the Duffing oscillator in the presence of single-photon loss and external excitation. 
Our devices are measured in a hanger geometry, where there are two modes propagating in the transmission line in opposite directions; we label the right and left going modes by "+" and "-", respectively.

The quantum Langevin equation (QLE) for the intra cavity field $a(t)$ is given by  
\begin{align}
    \dot{a}(t) &= \frac{1}{i \hbar} \left[H, a(t) \right] - \frac{\kappa_+ + \kappa_- + \gamma}{2} a(t) \\
    & + \sqrt{\kappa_+} a_{\mathrm{in},+} + \sqrt{\kappa_-} a_{\mathrm{in},-} + \sqrt{\gamma} a_{\mathrm{in},i}
\end{align}
Here, $a$ is the field operator associated with the intra-cavity field, while $a_{\mathrm{in},j}$ describe the field operators of the input fields associated with the three ports, the two input ports of the transmission line as well as the lossy environment. Accordingly, the first term in the QLE describes the deterministic evolution due to the Hamiltonian of the system, while the second term describes energy relaxation into the transmission line and internal degrees-of-freedom, for instance due to the presence of TLSs or non-equilibrium quasiparticles. 

The coupling rates to the two modes in the transmission line are $\kappa_+$ and $\kappa_-$, respectively, while the coupling rate to the intrinsic degrees of freedom is $\gamma$. In general, the two coupling rates to the left and right propagating modes in the transmission line can be different, giving rise to an asymmetric hanger geometry. However, for simplicity, we will assume a symmetric hanger geometry first.
\subsubsection*{Boundary conditions}
According to the input-output formalism\,\cite{Gardiner_Collett_1985}, there are boundary conditions relating the incident, the outgoing and the intra-cavity field for each of the three modes associated with the environment:
\begin{align}
    a_{\mathrm{out},+}(t) &= a_{\mathrm{in},+}(t) - \sqrt{\kappa_+} a(t) \\ 
    a_{\mathrm{out},-}(t) &= a_{\mathrm{in},-}(t) - \sqrt{\kappa_-} a(t) \\
    a_{\mathrm{out},i}(t) &= a_{\mathrm{in},i}(t) - \sqrt{\gamma} a(t)
\end{align}
While the first two equations describe the left- and right-traveling modes in the transmission line, respectively, the last equation describes uncontrolled degrees-of-freedom not associated with the input port. Notably, the sign convention of these boundary conditions can vary from source to source. 

\subsubsection*{Symmetric hanger geometry}
For a symmetric coupling to the transmission line, we can assume $\kappa_+ = \kappa_- = \kappa/2$, where $\kappa$ is the total decay rate into the transmission line. Moreover, if we drive the system only through one of the two ports, and assume that the environmental modes are cold ($a_{\mathrm{in},-} = a_{\mathrm{in},i} = 0$), we can simplify the QLE:
\begin{equation}
    \dot{a}(t) = -i\omega_0 a(t) - iK a^\dagger a^2 - \frac{\kappa + \gamma}{2} a(t) + \sqrt{\frac{\kappa}{2}} a_{\mathrm{in},+}. 
\end{equation}
\subsubsection*{Classical steady-state}
In order to solve for the classical steady-state of the QLE, a common ansatz for the field components of interest is to move into the rotating frame of the drive, and decompose the field amplitudes into the sum of a quantum part and a classical part.
\begin{align}
    a_{\mathrm{in},+}(t) &= \left[b_{\mathrm{in},+}(t)  + A_{\mathrm{in},+}(t) \right]e^{-i \omega t} \\ 
    a_{\mathrm{out},+}(t) &= \left[b_{\mathrm{out},+}(t) + A_{\mathrm{out},+}(t) \right]e^{-i \omega t} \\ 
    a(t) &= \left[b(t) + A(t) \right]e^{-i \omega t} 
\end{align}
Here, the $A$s represent the classical part of the field amplitude around which we evolve the quantum part represented by the $b$ operators.
Inserting the ansatz above into the QLE, we arrive at
\begin{equation}
     \dot{A} = -i\left(\omega_0 - \omega\right)A - iK A^* A^2 - \frac{\kappa + \gamma}{2} A + \sqrt{\frac{\kappa}{2}} A_{\mathrm{in},+}.   
\end{equation}
In the steady-state, the intra-cavity field is constant, and the time derivative vanishes.
\begin{equation}
    i\left(\omega_0 - \omega\right)A + iK A^* A^2 + \frac{\kappa + \gamma}{2} A = \sqrt{\frac{\kappa}{2}} A_{\mathrm{in},+}. 
\end{equation}
By multiplying the equation above by its complex conjugate, we finally arrive at an equation which describes the average, steady-state photon number in the Kerr oscillator for a given drive amplitude and detuning:
\begin{align}
    &\left[\left(\omega_0 - \omega\right)^2 + \left(\frac{\kappa + \gamma}{2}\right)^2 \right]|A|^2 \\ \notag
    &+ 2 K \left(\omega_0 - \omega\right) |A|^4 + K^2 |A|^6 = \frac{\kappa}{2} |A_{\mathrm{in},+}|^2. 
\end{align}
\subsubsection*{Scale-invariant steady state}
Following the example of Ref.~\cite{Eichler2014} and Ref.\,\cite{Anferov_2020}, we can re-express the equation of the steady-state photon number of the driven Duffing oscillator in a scale-invariant version:
\begin{equation}
    \left(\delta^2 + \frac{1}{4}\right) n - 2 \delta \xi n^2 + \xi^2 n^3 = \frac{1}{2}.
    \label{EQ_steady_state_invariant}
\end{equation}
Notably, for a Kerr oscillator measured in a single-port reflection configuration, the value on the right-hand side is 1 \,\cite{Eichler2014}. We have introduced the following scale-invariant parameters:
\begin{align}
    \delta &= \frac{\omega - \omega_0}{\kappa + \gamma}, \qquad n = \frac{|A|^2}{|\tilde{A}_{\mathrm{in}}|^2} \\ \notag 
    \tilde{A}_{\mathrm{in}} &= \frac{\sqrt{\kappa} A_{\mathrm{in}}}{\kappa + \gamma} \qquad \xi = \frac{K |\tilde{A}_{\mathrm{in}}|^2 }{\kappa + \gamma},
    \label{EQ_Kerr_scale_invariant_parameters}
\end{align}
Here, $\delta$ is the detuning normalized by the total resonator linewidth, $n$ is the scale-invariant photon number, $\tilde{A}_\mathrm{in}$ is the re-scaled input-field amplitude, and $\xi$ is the effective drive strength, which is related to the Kerr coefficient and the input field. 

In the scale-invariant formulation, the intra-cavity photon number develops multiple solutions above the critical drive strength $|\xi_{\mathrm{crit}}| = 2 / \sqrt{27}$, which is accompanied by an induced frequency shift $|\delta_{\mathrm{crit}}| = \sqrt{3} / 2$. Here, the sign of the frequency shift depends on the sign of the Kerr coefficient. For a negative Kerr coefficient, the induced frequency shift is negative, and for a positive Kerr coefficient the frequency shift is positive. In superconducting resonators, the Kerr coefficient typically arises from the finite critical current of the film, and is therefore negative ($\xi_{\mathrm{crit}} = -2 / \sqrt{27}$). 

The maximal value of the scale-invariant photon number is $\mathrm{max}(n) = 2$, independent of the value of $\xi$ and the total linewidth of the oscillator\,\cite{Anferov_2020}. Therefore, the average number of photons in the resonator $\bar{n}$, introduced in Eq.\,\eqref{nbar-via-P} in the main text, is expressed as 
\be \label{nbar-calculation-Appendix}
    \bar{n} = |A|^2 = \frac{2 \kappa}{(\kappa  + \gamma)^2} \frac{P_{\mathrm{in}}}{\hbar \omega_0}.
\ee
Here, we use the relation between the internal and external loss rate and the internal and external quality factor $\gamma = \omega_0 / Q_{\mathrm{i}}$ and $\kappa = \omega_0 / Q_{\mathrm{e}}$, respectively. 

\subsection{Analytical solutions}
\label{Sec_analytical_solutions}
The relation between the average intra-cavity photon number and the drive power is described by a cubic equation. For the implementation of a fast fitting algorithm, it is beneficial to derive analytical expressions for the solutions compared to a numerical approach. Since the solutions can be complex, we will keep only the real solutions. 

We can express Eq.\,\ref{EQ_steady_state_invariant}  in the general form
\begin{equation}
    n^3 + p_1 n^2 + p_2 n + p_3 = 0,
\end{equation}
where we have defined
\begin{equation}
    p_1 = - \frac{2 \delta }{\xi}, \qquad p_2 = \frac{\delta^2 + \frac{1}{4}}{\xi^2}, \qquad p_3 = -\frac{1}{2 \xi^2}.
\end{equation}
The system will have a single real solution below the point of bifurcation ($|\xi / \xi_{\mathrm{crit}}| < 1$ ) and there are three real solutions above. We can distinguish between the two cases by calculating the discriminant $M$
\begin{equation}
    M = \nu^2 - \eta^3,
\end{equation}
with 
\begin{equation}
    \nu = \frac{2 p_1^3 - 9p_1p_2 + 27p_3}{54} \quad \mathrm{and} \quad \eta = \frac{p_1^2 - 3p_2}{9}.
\end{equation}
If the discriminant is positive $M > 0$, the intra-cavity photon number $\bar{n}$ has a single real solution, while it has three solutions if the discriminant is negative $M < 0$.

\begin{figure}[t]
    \includegraphics[width =0.8\linewidth]{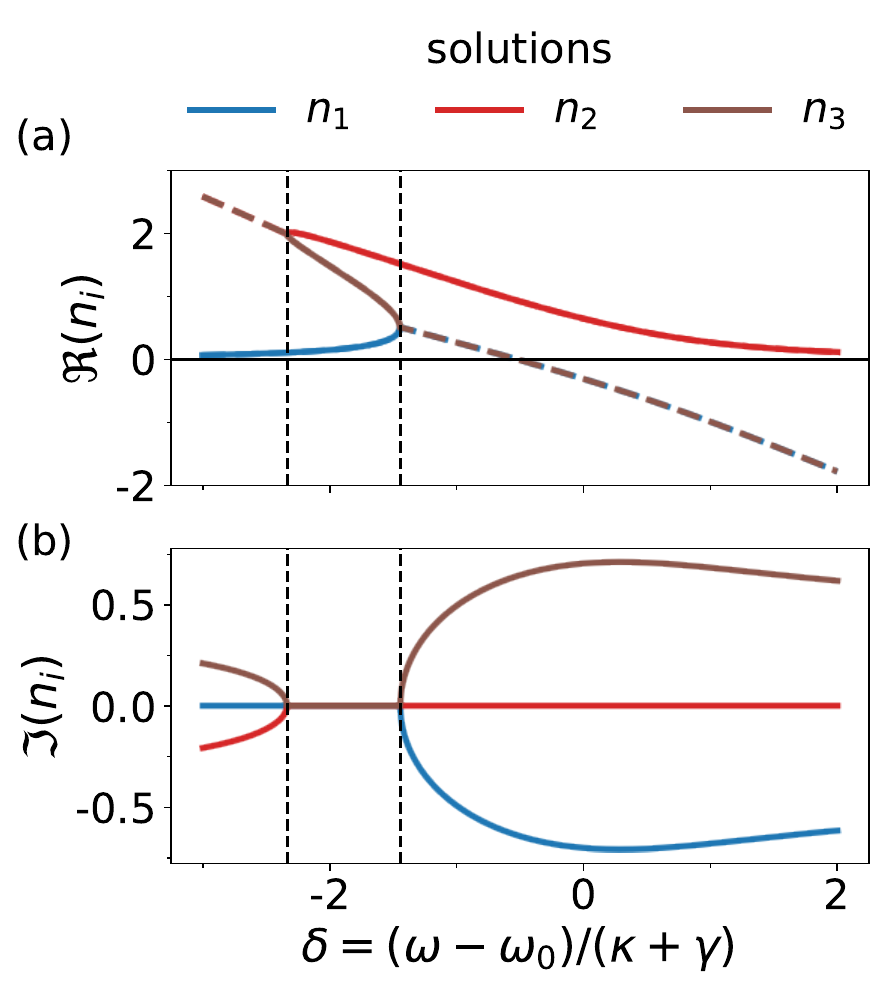}
    \caption{\textbf{Duffing oscillator: Steady state solutions of the scale-invariant photon number.} Real (panel a) and imaginary part (panel b) of the steady-state solutions $n_1, n_2, n_3$ of the scale-invariant mean photon number in a driven Kerr oscillator according to Eq.\,\ref{EQ_steady_state_invariant} as a function of detuning $\delta = (\omega - \omega_0) / (\kappa + \gamma)$. The effective drive strength $\xi / \xi_{\mathrm{crit}} = 3$ is well beyond the onset of bifurcation, resulting in a total of three solutions, indicated by the colors red, blue and brown. The dashed lines indicate the unstable (real) solutions for which the imaginary part is non-zero. The vertical dashed lines indicate the region in which multiple solutions exist at the same time, which is only the case for $|\xi / \xi_{\mathrm{crit}}| \geq 1$
}
    \label{Fig_Duffing_solutions}
\end{figure}

\subsubsection*{Single stable solution \underline{$M > 0$:}}
First, we start with the simpler case of a positive discriminant resulting in a single real solution. We can calculate the intra-cavity photon number as
\begin{equation}
    n = \chi + \zeta - \frac{p_1}{3}
\end{equation}
with
\begin{equation}
    \chi = \sqrt[3]{- \nu + \sqrt{M}} \quad \mathrm{and} \quad  \zeta = \sqrt[3]{- \nu - \sqrt{M}}
\end{equation}
The branch of the cubic root is to be taken as follows:
\begin{equation}
    \chi^\prime = - \mathrm{sign}(\nu)\left(|\nu| + \sqrt{M}\right)^{1/3}, \qquad \zeta = \frac{\eta}{\chi^\prime}
\end{equation}

\subsubsection*{Multiple stable solutions \underline{$M < 0$:}}
Second, we can look into the case of a negative discriminant, for which we find three solutions to the intra-cavity photon number:
\begin{align}
    n_1 &= - 2 \sqrt{\eta} \cos{\left(\frac{\theta}{3}\right)} - \frac{p_1}{3} \\ 
    n_2 &= - 2 \sqrt{\eta} \cos{\left(\frac{\theta + 2 \pi}{3}\right)} - \frac{p_1}{3} \\
    n_3 &= - 2 \sqrt{\eta} \cos{\left(\frac{\theta - 2 \pi}{3}\right)} - \frac{p_1}{3} 
\end{align}
with
\begin{equation}
    \theta = \arccos{\left(\frac{\nu}{\sqrt{\eta^3}}\right)}.
\end{equation}
The stable solutions are $n_1$ and $n_2$, while $n_3$ is unstable. Which stable solution we will find in the resonator depends on the direction of the frequency sweep when measuring the transmission coefficient. 

An example for the three solutions to Eq.\,\ref{EQ_steady_state_invariant} is shown in Fig.\,\ref{Fig_Duffing_solutions} for an effective drive strength $\xi / \xi_{\mathrm{crit}} = 3$ as a function of the detuning $\delta$. As indicated by the vertical dashed lines, there is a region of detuning for which we find three solutions with vanishing imaginary part, while in the other regions, only one solution has a vanishing imaginary part. 

\begin{figure}[t]
    \includegraphics[width =1\linewidth]{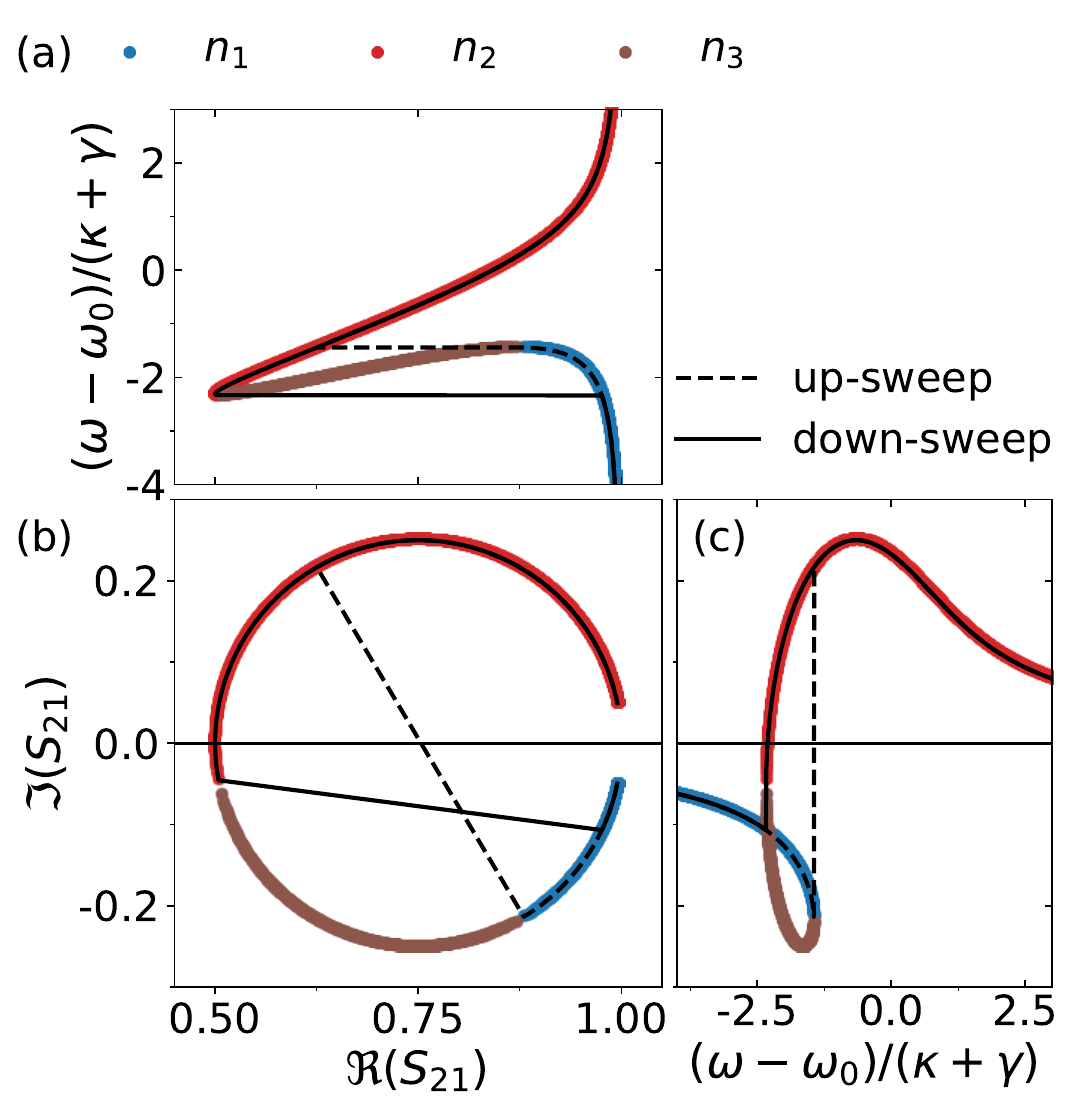}
    \caption{\textbf{Duffing oscillator: Transmission coefficient.} Simulated transmission coefficient $S_{21}$ as a function of detuning $\delta = (\omega - \omega_0) / (\kappa + \gamma)$ for an effective drive strength $\xi / \xi_{\mathrm{crit}} = 3$. The panels (a) - (c) show the frequency dependence of the real part, the transmission coefficient in the complex plane, and the frequency dependence of the imaginary part, respectively. The colors indicate the value of the transmission coefficient for the three solutions of the intra-cavity photon number (see Fig.\,\ref{Fig_Duffing_solutions}). The dashed, black line indicates the resulting transmission coefficient when sweeping the frequency in positive direction (up-sweep), while the solid, black line indicates the resulting response for a negative frequency sweep (down-sweep). The photon number $n_3$ is not reached as it is an unstable solution. 
}
    \label{Fig_transmission_coefficient}
\end{figure}

\begin{figure*}[t]
    \includegraphics[width =0.7\linewidth]{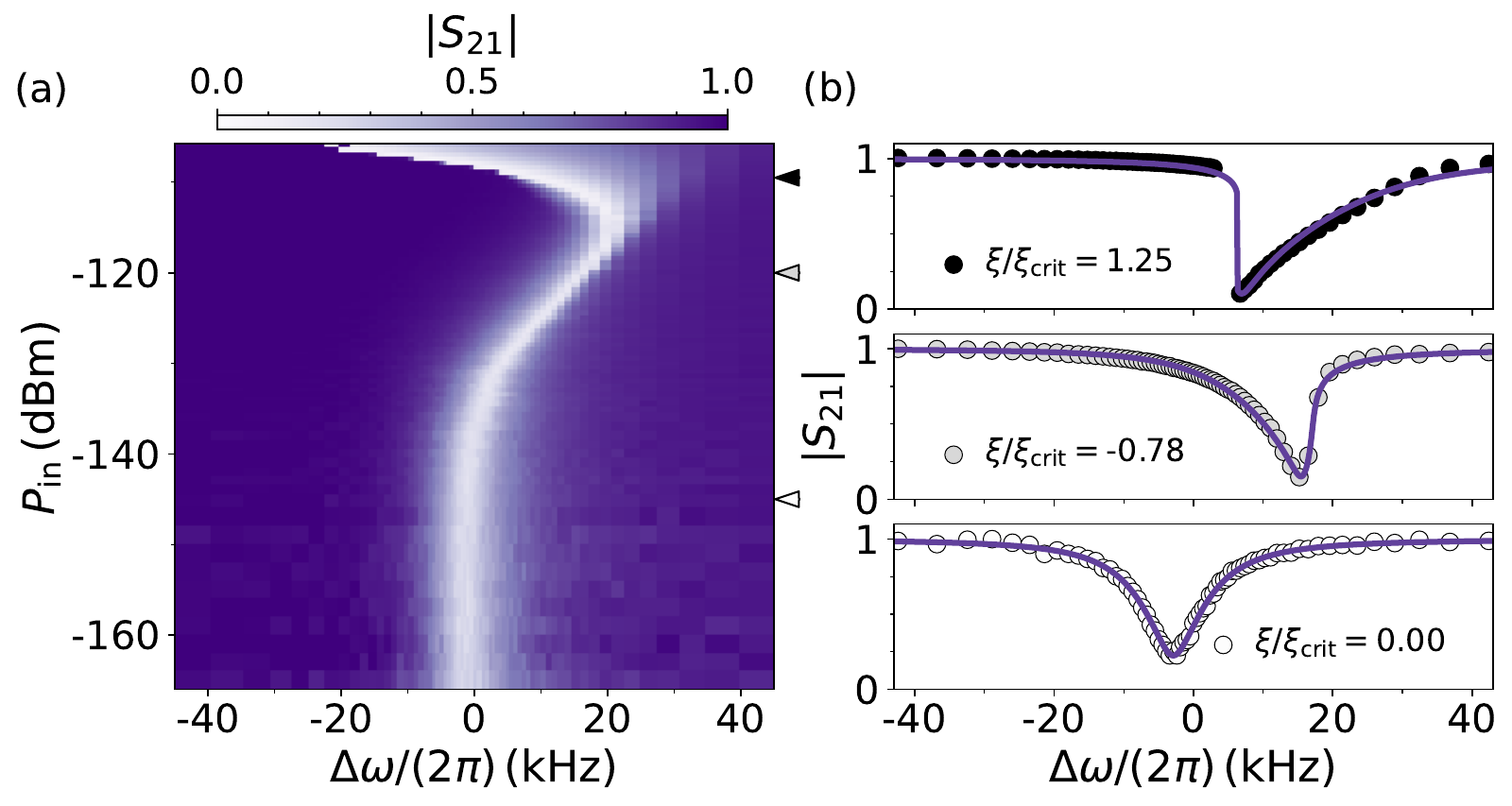}
    \caption{\textbf{Power dependence of the transmission coefficient:} (a) Absolute value of the transmission coefficient $|S_{21}|$ as a function of the readout frequency $\Delta \omega / (2\pi) = f - f_0$ and the readout power $P_{\mathrm{in}}$ measured around the resonance frequency $f_0$ of a lumped-element grAl resonator. With increasing readout power, the resonance first shifts in positive frequency direction, before it turns around and continues to shift in the negative direction. (b) Line-cuts of the magnitude $|S_{21}|$ for three different readout powers indicated by the similarly colored arrows in panel a. The solid line is a fit to the transmission coefficient of the semi-classical Duffing oscillator (Eq.\,\ref{Eq_Duffing_S21}), from which we extract the effective, scale-invariant drive strength $\xi$ normalized by the critical drive strength $\xi_{\mathrm{crit}} = - 2 / \sqrt{27}$ at which the resonator bifurcates. At low power (bottom panel), the shape is Lorentzian. At intermediate powers (central panel) close to the saturation of the positive frequency shift, the shape of the resonance is distinctly asymmetric with a sharp frequency dependence above the resonance, indicative of a positive Kerr coefficient. By further increasing the readout power, the resonance starts to shift again in the negative frequency direction with a distinct asymmetry below the resonance. 
}
    \label{Fig_scattering_coefficient}
\end{figure*}

\subsection{Transmission coefficient}
We can calculate the frequency dependence of the transmission coefficient of the Kerr oscillator, measured in the hanger geometry, from the ratio of the out-going to the incident field amplitude:
\begin{align}
    S_{21}(\omega)&= \frac{A_{\mathrm{out},+}(\omega)}{A_{\mathrm{in},+}(\omega)} \\
    &= 1 - \sqrt{\frac{\kappa}{2}} \frac{A}{A_{\mathrm{in},+}}
\end{align}
From the steady-state expression relating the intra-cavity field and the incident drive amplitude, we arrive at the transmission coefficient as a function of the detuning:
\begin{align}
    S_{21} &= 1 - \frac{\kappa}{\kappa + \gamma} \frac{1}{1 - 2i \left(\delta - \xi n \right)} \\
    &=1 - \frac{\kappa}{\kappa + \gamma - 2 i \left[\omega - (\omega_0 + K \bar{n}) \right]} 
\end{align}
While the scale-invariant version is practically useful because the effective drive strength $\xi$ can be expressed in units of the critical drive strength $\xi_{\mathrm{crit}}$, the second version directly shows how the effective frequency of the oscillator shifts by $K \bar{n}$. Notably, the average photon-number itself is a function of detuning between drive and resonance frequency, and needs to be calculated using Eq.\,\ref{EQ_steady_state_invariant} for every value of $\omega$ following the expressions summarized in Sec.\,\ref{Sec_analytical_solutions}.

For the derivation of the transmission coefficient, we have used the physical definition of the imaginary number. In electrical engineering, the imaginary number is $j = -i$. Since the Vector Network Analyzer (VNA) defines the phase of the scattering parameters following the electrical engineering convention, we can simply translate the definition of the transmission coefficient:
\begin{equation}
    S_{21} = 1 - \frac{\kappa}{\kappa + \gamma} \frac{1}{1 + 2j \left(\delta - \xi n \right)} \label{Eq_Duffing_S21} 
\end{equation}

\subsubsection*{Accounting for the transfer function}
In a realistic measurement setup, we need to account for additional contributions to the scattering coefficient caused by the complex transfer function of the readout lines. The presence of a finite impedance mismatch in the readout line can give rise to a rotation of the transmission coefficient around its center point, which we account for by introducing a complex phase $\phi_{\mathrm{e}}$ for the external coupling rate\,\cite{Khalil_2012, Megrant_2012}. Moreover, given the finite speed of light and the finite length of the cable, we account for an offset phase of the transmitted signal $\phi_0$ and a finite electrical delay $\tau$. Including these additional effects, the scattering coefficient reads
\begin{equation}
    S_{21} = \left(1 - \frac{\kappa e^{j \phi_{\mathrm{e}}}}{\kappa + \gamma} \frac{1}{1 + 2j \left(\delta - \xi n \right)}\right)e^{j \phi_0} e^{j \tau \omega}.
\end{equation}
For the fitting of every dataset, we remove these additional effects caused by the readout lines.

\begin{figure*}[t]
    \includegraphics[width =0.9\linewidth]{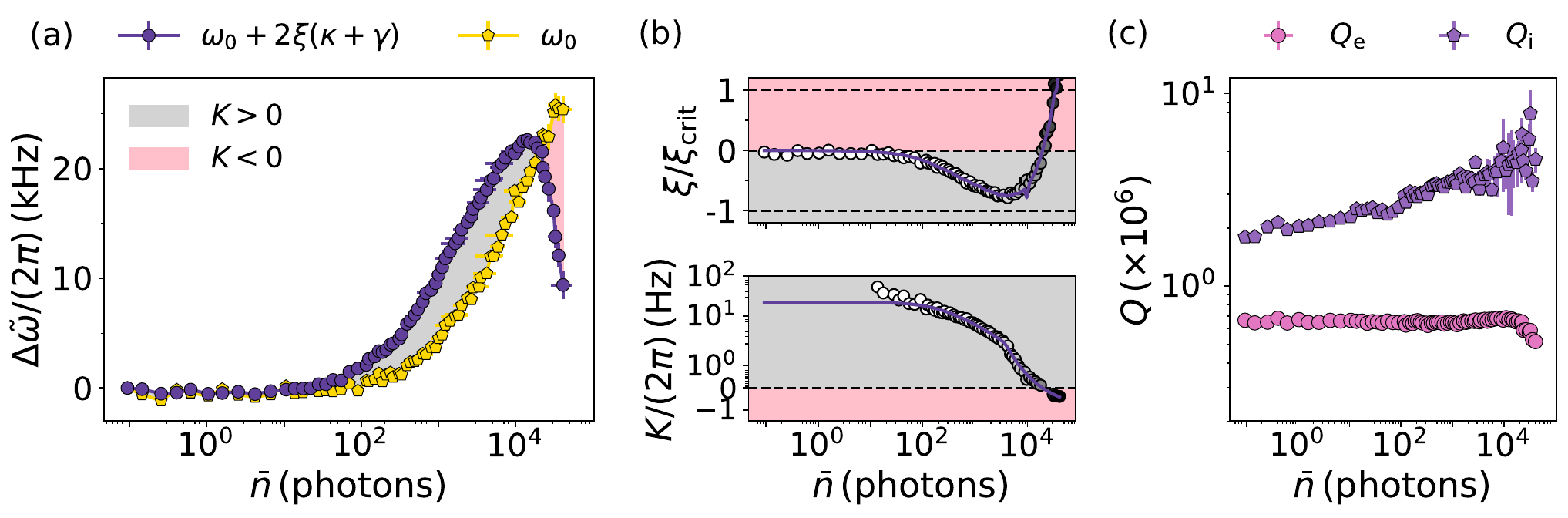}
    \caption{\textbf{
    Fitting of the $S_{21}(\omega)$ signal in the strongly nonlinear regime using the Duffing oscillator model.} 
    Parameters of the Duffing oscillator (frequency $\omega_0$, nonlinearity $K$, as well as $Q_{i}$ and $Q_e$) are fitted for each photon number $\bar{n}$ independently. (a) $\omega_0$ parameter (golden markers) and frequency with a non-linear correction, $\omega_0 + 2 \xi (\kappa + \gamma)  = \omega_0 + K\bar{n}$ (purple markers) as a function of $\bar{n}$. 
    Since $\omega_0$ deviates from the low-power value $2\pi f_0$ at $\bar{n} \gtrsim 10^2$, the Duffing oscillator alone cannot explain the measurements. The magnitude of the Kerr induced frequency shift is the difference between the golden markers and the purple markers. Consequently, the grey region indicates the photon number range in which the Kerr coefficient is seemingly positive, while it is seemingly negative in the pink region. (b) Fitted effective drive strength $\xi / \xi_{\mathrm{crit}}$ with $\xi_{\mathrm{crit}} = -2 / \sqrt{27}$ (top panel) and calculated Kerr coefficient (bottom panel) following Eq.\,\eqref{Eq_Kerr_xi} as a function of photon number. The effective drive strength unexpectedly changes sign. The solid purple line indicates a fit according to a phenomenological model of the Kerr coefficient Eq.\,\eqref{Eq_Kerr_phen} from which we can extract the values of $K_+ = 2\pi \times \left(23.1\pm3.0 \right)\,\mathrm{Hz}$ and $|K_-| = 2\pi \times \left(1.06\pm0.05 \right)\,\mathrm{Hz} $. (c) Internal and external quality factor as a function of photon number. While the external quality factor remains constant up to the largest power values, where the fit-result becomes less reliable, the internal quality factor monotonically increases with power.
}
    \label{Fig_Duffing_fit_parameters}
\end{figure*}

\section{Extracting the circuit parameters}
\label{Sec_extracting}
The frequency and power dependence of the absolute value of the transmission coefficient $S_{21}$ are shown Fig.\,\ref{Fig_scattering_coefficient} (panel a) for the device with $f_0 = 7.53\,\mathrm{GHz}$, together with three line-cuts taken at different readout powers (panel b) as indicated by the colored triangles. 

We fit the experimentally measured dependence of the complex transmission coefficient $S_{21}$ on the drive frequency $f$ to the Duffing oscillator formula in Eq.\,\ref{Eq_Duffing_S21} to obtain the relevant system parameters: the resonance frequency $\omega_0$, the internal and external loss rates $\gamma$ and $\kappa$, respectively, as well as the drive strength $\xi$. The internal and external quality factors are then obtained as
\be \label{app-Qi-Qe}
    \frac{1}{Q_{\mathrm{i}}} = \frac{\omega_0}{\gamma}, \qquad \frac{1}{Q_{\mathrm{e}}} = \frac{\omega_0}{\kappa} .
\ee
Note that this procedure works even at readout powers at which a linear circle fit would fail. The resulting value of the effective Kerr constant still has a dependence on $\bar{n}$ and changes sign from negative to positive values as $\bar{n}$ is increased. 

The fit results for the transmission coefficient are indicated in Fig.\,\ref{Fig_scattering_coefficient} (panel b) by the solid purple lines. Before we discuss the extracted loss rates and quality factors, we first discuss the signatures of the non-linear response.
At the lowest power values, the response of the resonator is fully symmetric (bottom panel) and the effective drive strength is $\xi / \xi_{\mathrm{crit}} = 0$. With increasing readout power, the asymmetry of the transmission coefficient becomes increasingly more pronounced towards positive detunings (central panel), reminiscent of a Duffing oscillator with positive Kerr coefficient and positive $\xi$. Correspondingly, the extracted drive strength $\xi / \xi_{\mathrm{crit}} = -0.78$ is negative, since we assume a negative critical drive strength $\xi_{\mathrm{crit}} = -2 / \sqrt{27}$. At even higher readout powers (top panel), we observe the bifurcation towards negative detunings, which is in agreement with the expected negative Kerr coefficient in superconducting resonators. Correspondingly, the extracted effective drive strength $\xi / \xi_{\mathrm{crit}} = 1.25$ is positive and larger than 1. The change in the sign of $\xi$ indicates a change of sign in the Kerr coefficient. Notably, the pronounced distortion of the transmission coefficient is only visible in our devices because they have low loss rates.

\subsubsection*{Power dependence of fit parameters}

The goal of this Section is to extend the data analysis to higher powers, $\bar{n}>4 \cdot 10^3$ up to the values close to the bifurcation point, far beyond the range considered in the main text. For that purpose, we use the Duffing oscillator model characterized by the frequency $\omega_0$ and nonlinearity $K$, see Eq.~\eqref{Duffing-Hamiltonian}. In the following, $\omega_0$ and $K$ are treated as fitting parameters for the measured $S_{21}(\omega)$ dependence at each readout power  $P_{\mathrm{in}}$. The parameter $K$ is not to be confused with the value of $K_0$ (obtained by a different procedure) in Secs.~\ref{sec:evaluating-Kerr} and \ref{sec:heat-balance-model} of the main text.

The extracted fit parameters are shown in Fig.\,\ref{Fig_Duffing_fit_parameters} as a function of the intra-cavity photon number $\bar{n}$, which is calculated via Eq.\,\eqref{nbar-via-P}. We plot the resonance frequency $\omega_{0}$ (golden markers) and the resonance frequency shifted by the non-linear Kerr effect $\tilde{\omega}_{0} = \omega_0 + 2 \xi (\kappa + \gamma)  = \omega_0 + K\bar{n}$ (purple markers) in panel (a). We subtract the low power value of $\omega_0$ for better visibility. The fit distinguishes between two effects: An unexpected shift in $\omega_0$ with increasing photon number, and the non-linear frequency shift caused by the Kerr effect.
We can observe the interplay between these two effects already in the raw data by comparing the lowest-power measurement (bottom panel in Fig.\,\ref{Fig_scattering_coefficient}) to the higher-power (central and top panel in Fig.\,\ref{Fig_scattering_coefficient}). According to the Duffing model, we expect a bifurcation at a critical detuning of $|\delta_{\mathrm{crit}}| = \sqrt{3}/2$, which for the device discussed, would be on the order of $12\,\mathrm{kHz}$. However, the positive frequency shift is significantly larger (central panel) despite the resonator bifurcating. Moreover, even though the highest-power measurement shows the characteristic behavior of a Duffing oscillator with negative Kerr coefficient, the resonance is still visibly shifted towards positive detunings compared to the low power value, indicative of an actual change in $\omega_0$. While this additional power dependence is not explicitly visible in the expression of the transmission coefficient, it is similar in spirit to the power dependence of the internal loss rate, which is also typically not explicitly expressed.

\begin{figure}[t]
    \includegraphics[width =\linewidth]{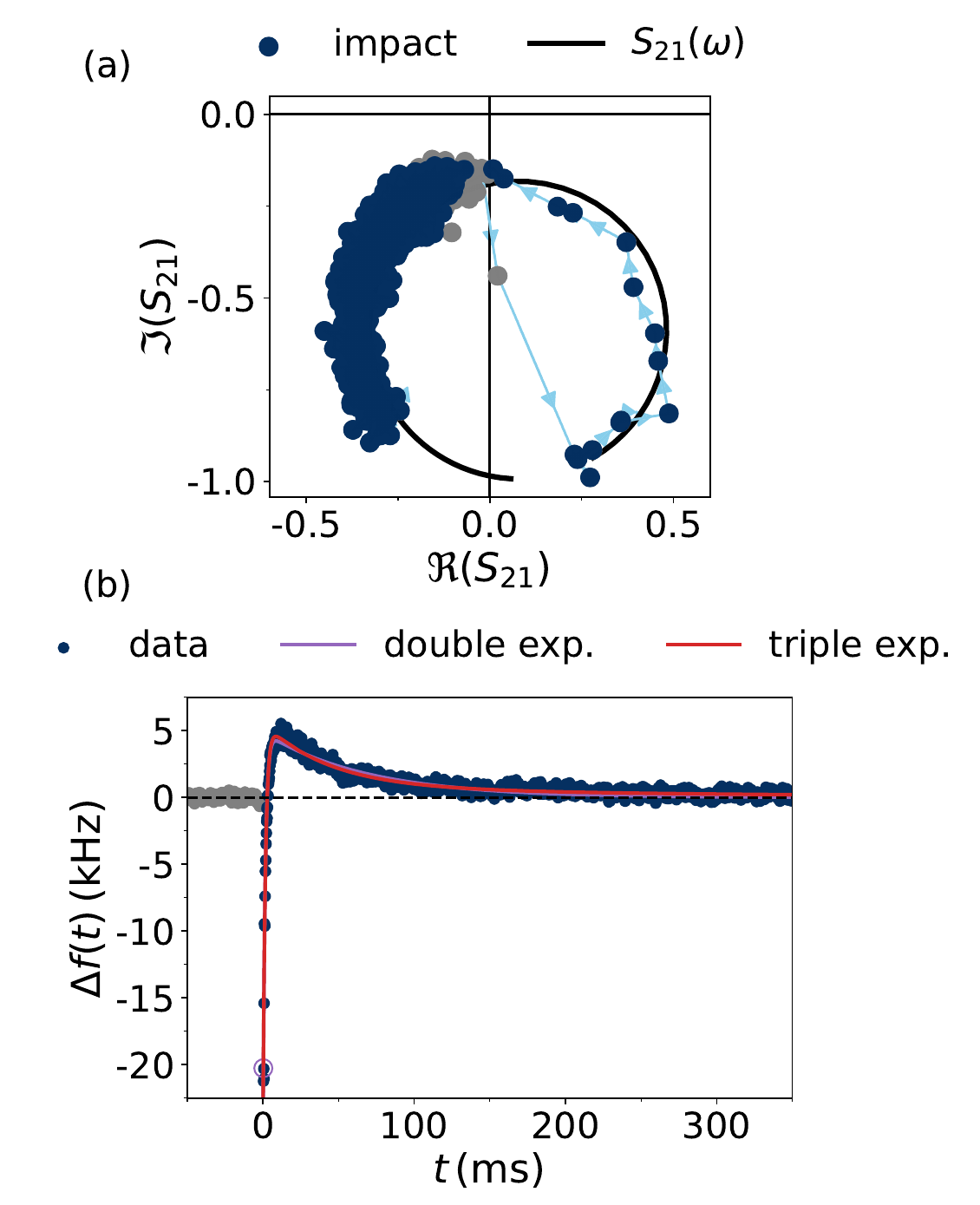}
    \caption{\textbf{Analysis of high energy events}. Resonance frequency: $\omega_0 / (2\pi) = 5.31\,\mathrm{GHz}$, Temperature: $T = 23\,\mathrm{mK}$, Cool-down: 11. (a) Time-dependence of the complex transmission coefficient $S_{21}(t)$, monitored at a fixed readout frequency $\omega_{\mathrm{r}}$ slightly below the steady-state resonance frequency before (grey markers) and during (blue markers) a high energy event. The IFBW is $5\,\mathrm{kHz}$ and the mean photon number is $\bar{n}\approx 10500\,\mathrm{photons}$. The arrow of time is indicated by the sky-blue lines. The change in the transmission coefficient is caused by an underlying change of the resonance frequency. The frequency dependence of the complex transmission coefficient $S_{21}(\omega)$ (black solid line) is used as a look-up table to convert the signal $S_{21}(t)$ into a time-dependent resonance frequency $\omega_0(t)$. (b) Time-dependence of the resonance frequency with respect to the steady-state value. The energy deposited by the high-energy event results in the creation of quasiparticles, shifting the resonance frequency downwards, followed by a rapid relaxation process which overshoots and slowly relaxes back to equilibrium. We use a double (purple solid line) and a triple (red solid line) exponential fit to approximate the complex relaxation process (see Fig.\,\ref{Fig_high_energy_events_5p31GHz_statistics}).
}
    \label{Fig_high_energy_events_5p31GHz}
\end{figure}

\begin{figure*}[t!]
    \includegraphics[width =0.9\linewidth]{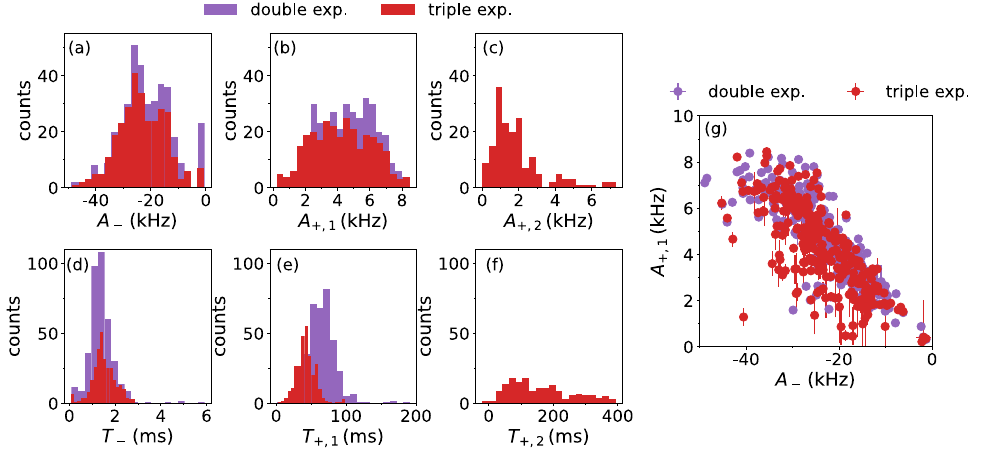}
    \caption{\textbf{Statistics of high energy events.} Device frequency: $\omega_0 / (2\pi) = 5.31\,\mathrm{GHz}$, Temperature: $T = 23\,\mathrm{mK}$, Cool-down: 11. From the double and triple exponential fit, shown in Fig.\,\ref{Fig_high_energy_events_5p31GHz}, we extract the magnitude of the frequency shift in negative and positive direction $A_-, A_{+,1}$ and $A_{+,2}$ (only for triple exp.), respectively, shown in panels (a) - (c), together with the exponential timescales of the relaxation process $T_-, T_{+,1}$ and $T_{+,2}$ (only for triple exp.), shown in panels (d) - (f). The results for both fit functions are consistent with each other for the parameters present in both models, with the main difference in the timescale of the initial, rapid decay of the positive frequency shift shown in panel (e). (g) Correlation between the magnitude of the initial negative frequency shift and the following positive frequency shift. The observed correlation is consistent with the picture of a temporary increase of the substrate temperature caused by the high-energy event. An event which causes a large negative frequency shift also gives rise to a large positive frequency shift.
}
    \label{Fig_high_energy_events_5p31GHz_statistics}
\end{figure*}

The effective drive strength is plotted in the upper panel of Fig.\,\ref{Fig_Duffing_fit_parameters}(b) and shows the aforementioned sign change. While the scale-invariant parameters are very useful when comparing the behavior of different samples, we can convert the effective drive strength into a Kerr coefficient. Using the relation between the incident photon number per unit time and the drive power $P_{\mathrm{in}} = \hbar \omega |\alpha_{\mathrm{in}}|^2$, we can calculate the power dependence of the effective drive strength
\begin{equation}
    \xi = \frac{K \kappa P_{\mathrm{in}}}{\hbar \omega \left(\kappa + \gamma\right)^3}.
\end{equation}
Alternatively, we can use the definition for the mean intra-cavity photon number $\bar{n} = |\alpha|^2$ and its relation to the input power according to Eq.\,\ref{nbar-via-P}
\begin{equation}
    \xi = \frac{K \bar{n}}{2 \left(\kappa + \gamma\right)}.
\end{equation}
For a device with constant Kerr coefficient and linewidth, we would expect a linear increase of the effective drive strength with $\bar{n}$. Instead, we observe a non-linear dependence of $\xi$ on the photon number and even a sign change. Since the linewidth changes only marginally in the investigated power range, we assign the change in $\xi$ to a change in the Kerr coefficient:
\begin{equation}
    K = \frac{2 \left(\kappa + \gamma\right)\xi}{\bar{n}}.
    \label{Eq_Kerr_xi}
\end{equation}
The numerator is the induced frequency shift, such that we recover the definition of the Kerr coefficient as a frequency shift per photon. The Kerr coefficient calculated from the fitted values of $\xi$ is shown in Fig.\,\ref{Fig_Duffing_fit_parameters} panel b (bottom). At low photon numbers, the frequency shift per photon is positive, while it is negative for photon numbers beyond $\bar{n} >40 000$. 

While we don't have an explanation for the sign change in $K$, we introduce a phenomenological model to capture the photon number dependence of $K$:
\begin{equation}
    \tilde{K}(\bar{n}) = -|K_-| + \frac{K_+}{\sqrt{1 + \left(\frac{\bar{n}}{\bar{n}_{\mathrm{c}}}\right)^\lambda}}.
    \label{Eq_Kerr_phen}
\end{equation}
Here, $K_-$ and $K_+$ are the fitting parameters, $n_{\mathrm{c}}$ is the critical photon number at which the positive frequency shift starts to saturate and $\lambda$ is the exponent. 
From the fit to our data, shown as the solid purple line in Fig.\,\ref{Fig_Duffing_fit_parameters} panel b, we extract $|K_-| = 2\pi \times \left(1.06\pm0.05 \right)\,\mathrm{Hz} $, $K_+ = 2\pi \times \left(23.1\pm3.0 \right)\,\mathrm{Hz}$, $n_{\mathrm{c}} = \left(180\pm50 \right)\,\mathrm{photons} $ and $\lambda = 1.32\pm0.04$. As shown in Fig.\,\ref{fig:Pdependence-fit}, the positive frequency shift with power disappears with increasing device temperature. 

The power dependence of the internal and external loss rates is shown in Fig.\,\ref{Fig_Duffing_fit_parameters}, panel (c), after converting them into the internal and external quality factors. While the external quality factor remains unchanged up to the highest readout powers, the internal quality factor monotonically increases. The change in the external quality factor is an artifact of the fitting procedure due to the pronounced non-linearity of the device at these elevated readout powers.

\section{Impact analysis}
\label{Sec_Impacts}
In this section, we describe the analysis procedure for obtaining the characteristic timescales of the relaxation process after a high-energy event (cf. Fig.\,\ref{Fig_impacts}). For the detection of these events, we continuously monitor the complex transmission coefficient $S_{21}(t)$ with a VNA at a fixed readout frequency $f_{\mathrm{r}}$ in close frequency vicinity to the resonance frequency $f_0$ of a device. A temporary change in the resonance frequency changes the detuning to the readout tone, resulting in a change in the complex scattering coefficient~\cite{Day2003}. By comparing the instantaneous response $S_{21}(t)$ to a high-resolution look-up table derived from the steady-state response $S_{21}(\omega)$, we can extract the time dynamics of the resonance frequency $f_0(t)$\,\cite{IoanPop-2018, grAl-inductors}. From these time traces, we identify the occurrence of impacts in post-processing by thresholding the instantaneous frequency shift, and use it to offset the time axis for every event to facilitate the analysis.

\subsection*{Device at 5.31 GHz}
Figure\,\ref{Fig_high_energy_events_5p31GHz} shows the typical instantaneous response of the resonator with $\omega_0 / (2\pi) = 5.31\,\mathrm{GHz}$ in the complex plane (colored markers) together with the frequency dependence of the steady-state transmission coefficient $S_{21}(\omega)$ (black solid line). The readout frequency is slightly below the steady-state resonance frequency and the estimated photon number is $\bar{n}\approx 10600\,\mathrm{photons}$. The intermediate frequency bandwidth (IFBW) of the measurement is $5\,\mathrm{kHz}$, resulting in a time resolution of around $\Delta t = 200\,\mathrm{\mu s}$. The blue markers indicate the response of the resonator right after a high energy event. First, the resonance frequency shifts downwards with respect to the readout tone presumably due to the creation of non-equilibrium quasiparticles, resulting in an almost instantaneous change of $S_{21}$, followed by a fast relaxation overshooting towards positive frequencies. The relaxation of the positive frequency shift is significantly slower. Very similar dynamics are observed in the devices with resonance frequencies $5.70\,\mathrm{GHz}$, $6.04\,\mathrm{GHz}$, $7.29\,\mathrm{GHz}$. 

The instantaneous change in the resonance frequency $\Delta f (t)$ is extracted from the data in the complex plane through the comparison to the look-up table $S_{21}(\omega)$ (black solid line) and shown in the panel (b). We extract the magnitude of the induced frequency shifts and the timescales of the relaxation process using a double or a triple exponential fit
\begin{equation}
    f(t) = A_{-} e^{- (t - t_0) / T_{-}} + A_{+,1} e^{- (t - t_0) / T_{+,1}}
\end{equation}
and
\begin{equation}
    f(t) = A_{-} e^{- \frac{(t - t_0)}{T_-}} + A_{+,1} e^{- \frac{(t - t_0)}{T_{+,1}}} + A_{+,2} e^{- \frac{(t - t_0)}{T_{+,2}}}
\end{equation}
Here, $t_0$ is the moment of impact. 

Overall, the triple exponential seems to provide a better fit: for some of the events, the relaxation of the positive frequency shift does not follow a simple single exponential decay. The statistics of the fit results are shown in Fig.\,\ref{Fig_high_energy_events_5p31GHz_statistics} for a total of 308 analyzed events measured within the course of 12 hours of measurement time. We observe variations in the magnitude of the initial negative frequency shift as well as the following overshoot in the positive direction. While the relaxation time of the negative frequency shift is rather constant, on the order of $T_- = 1-2\,\mathrm{ms}$, we observe larger variations in the relaxation time of the positive frequency shift, which is on the order of $50\,\mathrm{ms}$ in this device.

In the right panel of Fig. \ref{Fig_high_energy_events_5p31GHz_statistics} we plot the correlation between the magnitude of the negative and the positive frequency shift. The observed correlation is consistent with a temporary increase of the phonon temperature caused by the high-energy event. The different timescales could be caused by the different energy scales of the processes. During the initial negative frequency shift, the density of excess quasiparticles is enhanced by phonons with energies $\hbar \omega_{\mathrm{ph}} > 2 \Delta$. For the observation of the positive frequency shift, a much lower phonon energy is sufficient, as shown in Fig.\,\ref{Fig_temperature_dependence}. 

\begin{figure}[t]
    \includegraphics[width =0.8\linewidth]{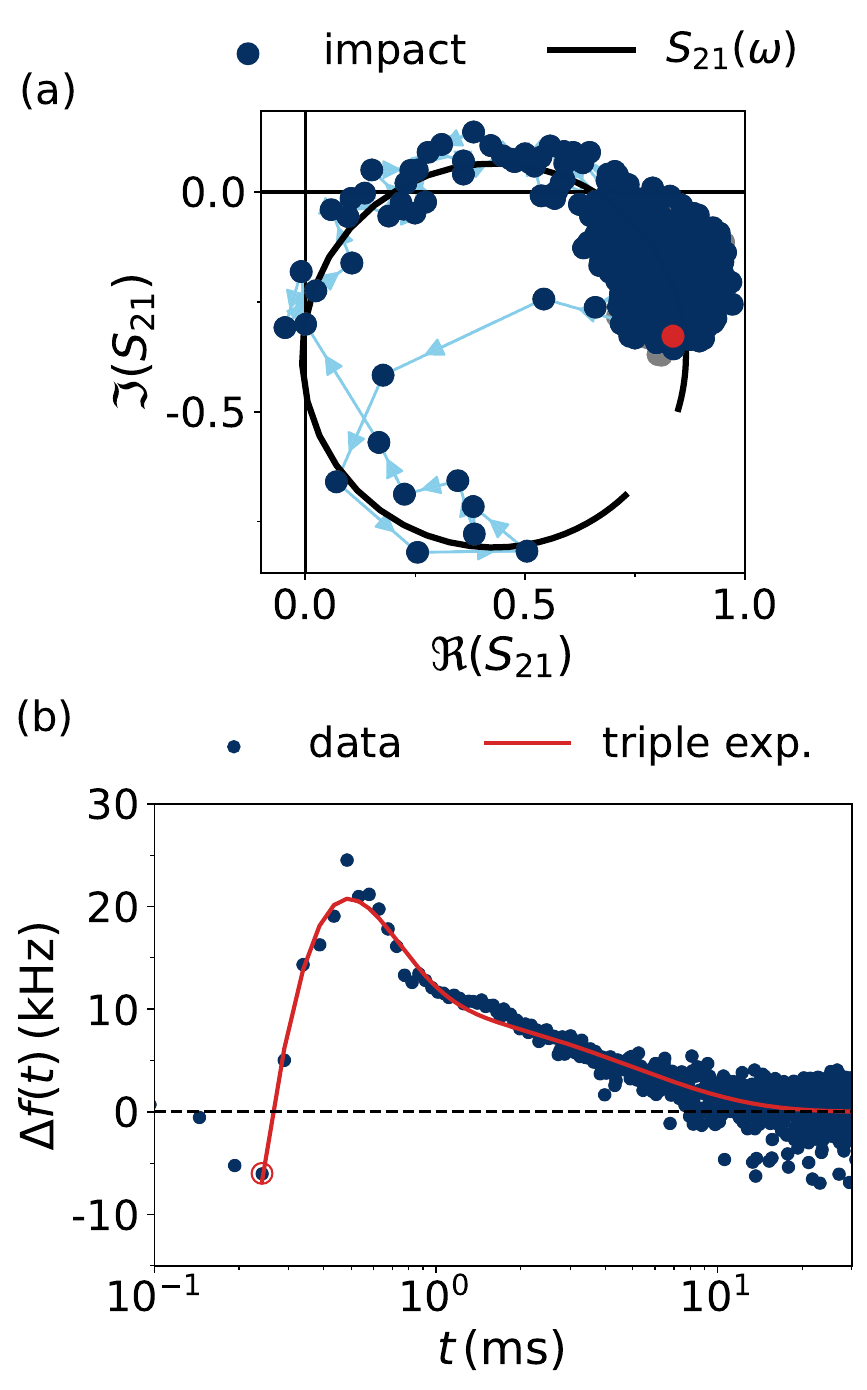}
    \caption{\textbf{Analysis of high energy events}. Resonance frequency: $\omega_0 / (2\pi) = 7.53\,\mathrm{GHz}$, Temperature: $T = 26\,\mathrm{mK}$, Cool-down: 15. (a) Time-dependence of the complex transmission coefficient $S_{21}(t)$, monitored at a fixed readout frequency $\omega_{\mathrm{r}}$ slightly above the steady-state resonance frequency before (grey markers) and during (blue markers) a high energy event. The IFBW is $20\,\mathrm{kHz}$ and the mean photon number is $\bar{n}\approx 15000\,\mathrm{photons}$. The arrow of time is indicated by the sky-blue lines. The change in $S_{21}$ is caused by an underlying change of the resonance frequency.  The frequency dependence of the complex transmission coefficient $S_{21}(\omega)$ (black solid line) is used as a look-up table to convert the signal $S_{21}(t)$ into a time-dependent resonance frequency $\omega_0(t)$. (b) Time-dependence of the resonance frequency with respect to the steady-state value. In contrast to the device shown in Fig.\,\ref{Fig_high_energy_events_5p31GHz}, the resonance frequency only marginally shifts downwards right after the impact; it then immediately shifts in positive direction and relaxes back into equilibrium at two distinct time scales. We use a triple (red solid line) exponential fit to approximate the complex relaxation process. The red marker indicates the starting point of the fit. We plot the time axis in log scale to highlight the fast dynamics right after the high energy event occurs. The time axis is offset to accommodate for the log scale.
}
    \label{Fig_high_energy_events_7p53GHz}
\end{figure}

\subsection*{Device at 7.53 GHz}
Figure\,\ref{Fig_high_energy_events_7p53GHz} shows the typical instantaneous response of the resonator with $\omega_0 / (2\pi) = 7.53\,\mathrm{GHz}$ in the complex plane. The readout frequency is above the steady-state resonance frequency, which makes the measurement more susceptible to positive frequency shifts of the resonator. However, we also performed measurements with a readout frequency below the resonator and did not observe a significant negative frequency shift. The estimated photon number is $\bar{n}\approx 15000\,\mathrm{photons}$. The intermediate frequency bandwidth (IFBW) of the measurement is $20\,\mathrm{kHz}$, resulting in a time resolution of around $\Delta t = 50\,\mathrm{\mu s}$. 

In contrast to the device shown in Fig.\,\ref{Fig_high_energy_events_5p31GHz}, the device does not show a pronounced negative frequency shift right after a high energy event. While we do record a negative shift, it is within the noise of the measurement. However, we observe a finite rise time of the positive frequency shift. After reaching the maximal positive frequency shift around $0.5\,\mathrm{ms}$, the instantaneous positive frequency shift seems to relax at two distinct timescales. Both timescales are significantly faster than what we observe in the other devices measured in CD 11, with $T_{+,1} = 0.22\pm0.1\,\mathrm{ms}$ and $T_{+,2} = 5.0\pm0.3\,\mathrm{ms}$. Notably, due to the more complex time evolution in this device and smaller signal-to-noise ratio, it is more difficult to identify the exact moment an impact occurs, resulting in a time offset which is subject to a certain degree of uncertainty. We measured a similar behavior in two other devices from the same batch.
\subsection*{Discussion}
The timescales of the relaxation process seem to be independent of the strip length. While we don't have an explanation for the faster relaxation in the devices measured in CD 15, we want to highlight that the main difference from other devices measured in CD 11 is the lower normal-state resistivity, which differs by a factor of 3-4, and the sapphire substrate type, which is HEM for the low-resistivity devices and EFG for the higher resistivity devices (see Tbl.\,\ref{table_info}).

\newpage

\bibliography{anomalous_grAl.bib}
\end{document}